\documentclass[fleqn,usenatbib]{mnras}

\usepackage{newtxtext,newtxmath}

\usepackage[T1]{fontenc}

\DeclareRobustCommand{\VAN}[3]{#2}
\let\VANthebibliography\thebibliography
\def\thebibliography{\DeclareRobustCommand{\VAN}[3]{##3}\VANthebibliography}

\usepackage{graphicx}	
\usepackage{amsmath}	
\usepackage{stackengine}
\usepackage{ragged2e}
\usepackage{comment}
\usepackage{soul}
\usepackage{lscape}
\usepackage{amssymb}
\usepackage[dvipsnames]{xcolor}
\usepackage[font=small,skip=0pt]{caption}
\usepackage[utf8]{inputenc}

\DeclareUnicodeCharacter{2500}{\mbox{\vrule height2.2ptdepth-1.8ptwidth.5em}}

\newcommand\angdot[1][\circ]{%
  \stackengine{0pt}{.}{${}^{\mathrm{#1}}$}{O}{l}{F}{F}{L}}
\newcommand\fermi{{\it Fermi}\,}

\title[The Dusty Blue BOAT]{JWST's Dusty Blue BOAT - GRB 221009A}

\author[Khang et al.]{Nguyen M. Khang$^{1}$\thanks{E-mail: m.k.nguyen@2023.ljmu.ac.uk (NMK)},
Gavin P. Lamb$^{1}$\thanks{E-mail: g.p.lamb@ljmu.ac.uk (GPL)}, 
Helena-M.S. Grabham$^{1}$,
Conor M. B. Omand$^{1}$,
Hamid Hamidani$^{2,3}$,\newauthor
Andrew J. Levan$^{4,5}$,
Nial R. Tanvir$^{6}$,
Valerio D'Elia$^{7}$,
Luca Izzo$^{8,9}$
\\
$^{1}$Astrophysics Research Institute, Liverpool John Moores University, IC2 Liverpool Science Park, 146 Brownlow Hill, Liverpool, L3 5RF, UK\\
$^{2}$School of Systems Information Science, Future University Hakodate, Hakodate, Hokkaido 041-8655, Japan\\
$^{3}$Astronomical Institute, Graduate School of Science, Tohoku University, Sendai 980-8578, Japan\\
$^{4}$Department of Astrophysics/IMAPP, Radboud University, 6525 AJ Nijmegen, The Netherlands\\
$^{5}$Department of Physics, University of Warwick, Coventry, CV4 7AL, UK\\ 
$^{6}$School of Physics and Astronomy, University of Leicester, University Road, Leicester, LE1 7RH, UK\\
$^{7}$Space Science Data Center (SSDC) ─ Agenzia Spaziale Italiana (ASI), I-00133 Roma, Italy\\
$^{8}$DARK, Niels Bohr Institute, University of Copenhagen, Jagtvej
155A, 2200 Copenhagen, Denmark\\
$^{9}$INAF, Osservatorio Astronomico di Capodimonte, Salita Moiariello
16, I-80121 Naples, Italy
}

\date{Accepted XXX. Received YYY; in original form ZZZ}

\pubyear{\the\year{}}

\begin{document}
\label{firstpage}
\pagerange{\pageref{firstpage}--\pageref{lastpage}}
\maketitle

\begin{abstract}
GRB 221009A, the Brightest Of All Time (BOAT), presents a challenge for afterglow modelling due to its low Galactic latitude and consequent high line-of-sight extinction. 
This has led to a wide range of conflicting values for the optical spectral index and dust extinction in the literature. 
We present a re-analysis of the afterglow spectra, using VLT X-Shooter data at 0.5, 4, and 10.5 days, and JWST NIRSpec/MIRI data at 13.3 days post-burst. 
We fit the data with single and smoothly broken power-law models and perform a joint fit with a double smoothly broken power-law (DSBPL) across all epochs. 
Our analysis reveals a strong degeneracy between the assumed extinction and the inferred intrinsic spectral index, particularly in the optical, explaining the diversity of previous results. 
The joint DSBPL fit yields a total line-of-sight extinction of $A_{V} = 4.40 \pm 0.01$ and a blue continuum, with an intrinsic spectral index of $\beta = 0.447 \pm 0.001$. 
Although marginally preferred by the spectral fits, a wind medium can be rejected by the temporal evolution of the afterglow light curve.
The fit spectral index and temporal decline are only consistent with a uniform density medium if an early jet break at $\sim$0.5–1.0 days is invoked. 
Our results imply a hard electron distribution index of $p = 1.89 < 2$, challenging standard particle acceleration models and suggesting a narrow, energetic jet core dominates the early optical-to-X-ray emission.
\end{abstract}




\section{Introduction}
As one of the most luminous electromagnetic (EM) events in the Universe, Gamma-ray Bursts (GRBs) are immensely powerful explosions lasting from seconds to minutes. Followed by a late-time afterglow, the entire process is capable of producing radiation over the whole EM spectrum from radio to gamma ($\gamma$)-rays. 
Since their discovery, space-based instruments have detected thousands of GRBs over nearly five decades, at a rate of a few events per day.

GRB 221009A was observed on the 9th October 2022 at 13:16:59 UT, and is amongst the best studied GRB events to date.
Famed for its extreme levels of brightness, GRB 221009A has been named the Brightest Of All Time (BOAT) GRB \citep{Burns_2023} and produced a record isotropic equivalent energy of up to $10^{55}$ erg \citep{lesage2023}.
In addition to possessing the highest-ever recorded peak flux and fluence, the BOAT was a relatively nearby event with a redshift of 0.151 \citep{malesani2025}, meaning that it was even more of an extreme outlier in terms of its observed radiative flux. 
So much so, that it even caused disturbances in the Earth's ionosphere \citep{Hayes_2022}.

The prompt emission of the BOAT lasted roughly 10 minutes, making it a long-duration GRB, and was detected by the \textit{Fermi} Gamma-ray Burst Monitor (GBM) \citep{Burns_2023}, the Neil Gehrels \textit{Swift}'s Burst Alert Telescope (BAT) and X-ray Telescope (XRT)
\citep{williams_maia2023}, \textit{Konus-Wind} \citep{Frederiks_2023}, and multiple other high energy observatories. 
An intensive afterglow follow-up of the BOAT was conducted over the following weeks, with the first-ever observations of a GRB afterglow obtained by the James Webb Space Telescope (JWST), including mid-infrared spectroscopy, alongside observations from the Hubble Space Telescope (HST), as well as multiple ground-based facilities.

Unfortunately, data for the BOAT, particularly in optical, infra-red (IR) \citep[e.g.,][]{kann2023, levan2023, laskar2023} and X-ray bands \citep[e.g.,][]{williams_maia2023, klinger2024} have been difficult to interpret
due to its sky position, with a line-of-sight through the Galactic plane of the Milky Way (MW).
Observations of the BOAT therefore suffer from high line-of-sight extinction ($A_V \approx 4.2$ from \citet{levan2023}), adding additional uncertainty to the intrinsic optical magnitudes and X-ray flux.
Particularly at X-ray, reflection from dust created a series of ring structures (dust echoes) that complicates the intrinsic light curve determination \citep{tiengo2023, vasilopoulos2023, zhao2024}, although, these rings have been discussed as useful probes of otherwise unobservable dust structures within the Milky Way \citep{Campana_2024,vaia2025, sneppen2025}. Still, the exact extent of the line-of-sight extinction is uncertain and the adopted value varies across the published literature (see Table\,\ref{tab:refs} for a sample of the literature values).
A detailed discussion of the extinction is given by \citet{kann2023}, who argue the reddening is best described as $E(B-V) = 1.69\pm0.03$ as opposed to the \citet{schlafly2011} value of $E(B-V) = 1.32$ for the sky location.

Despite the obscured viewpoint of the BOAT, the relatively low redshift combined with 
the extreme intrinsic luminosity
made its afterglow an attractive target for early observations. 
More importantly, near and mid-infrared (NIR/MIR) ranges are required to search for signatures of $r$-process nucleosynthesis within core-collapse GRB supernova \citep[see e.g.,][]{siegel2019, rastinejad2024}.
JWST took early observations of the counterpart at $\sim13$\,days post-burst;
these observations revealed a single power-law spectra consistent with non-thermal emission from the afterglow.
However, the spectra was bluer than expected with a $F_\nu \propto \nu^{-\beta}$ spectral index $\beta \sim 0.4$, which is below the typical assumption for a GRB afterglow of $0.5 < \beta \lesssim 1.0$ at optical/NIR \citep[e.g.,][]{sari1998}.
No supernova (SN) contribution was identified within these early observations  \citep{levan2023}.
Evidence of a low-luminosity SN was subsequently identified by JWST spectra in observations taken at much later times \citep[see][who confirmed the $\beta\sim0.4$ of the day-13 spectra]{blanchard2024}.

The JWST spectra has a supremely high signal-to-noise for GRB afterglow observations, and the blue single power-law structure of the intrinsic afterglow emission is reproduced/confirmed (although with marginally different posterior distributions).
However, the optical/NIR spectral energy distribution of $\beta \sim 0.4$ has been largely ignored when modelling the afterglow. 
This has resulted in multiple afterglow model fits that fail to reproduce this spectral data.
When considered, the afterglow models can become complex \citep[see,][where the authors ensure the ambient medium density profile is consistent with a decreasing cooling frequency -- required by the JWST spectra]{oconnor2023}.
As such, modelling of the BOAT's afterglow 
was inconclusive, resulting in a broad range of estimates for the spectral index at optical/NIR, the distribution of accelerated electrons, the ambient medium density structure, and the line-of-sight dust extinction and reddening.
Here we present a detailed analysis of the afterglow data of the BOAT, 
including the ground based spectra taken within the first 14 days by X-Shooter \citep{malesani2025}, and the JWST observations \citep{levan2023},
with the aim of re-evaluating the early afterglow spectra and understanding its physical nature.

In Section\,\ref{sec:methods}, we present the methods used to initially analyse the spectroscopic data for the first 14 days. 
In Section\,\ref{sec:results}, we show the results of the analysis described in the previous section.
Our results are discussed and additional analysis performed in Section\,\ref{sec:discussion}.
Finally, our conclusions are presented in Section\,\ref{sec:conclusions}.

\section{Methods}\label{sec:methods}
The BOAT datasets that are used for this analysis came from the following instruments.
\citet{malesani2025} obtained spectroscopic observations via the European Southern Observatory Very Large Telescope (ESO VLT Unit 3, Melipal) using the X-Shooter spectrograph.
The wavelength range of the X-Shooter spectra is $0.3 \leq \lambda \leq 2.5$\,$\mu$m.
\citet{levan2023} obtained spectroscopy with Near Infrared Spectrograph (NIRSPEC) and the Mid Infrared Instrument (MIRI), in the combined range $0.6 \leq \lambda \leq 13.1$\,$\mu$m, on board JWST $\sim13$\,days after the burst trigger.


\subsection{Data}

We mask the X-Shooter data in the wavelength ranges $\lambda<0.325$, $1.34 \leq \lambda \leq 1.46$, $1.78 \leq \lambda \leq 1.95$, and $\lambda>2.45$\,$\mu$m, and for MIRI at wavelengths $\lambda>8.80$\,$\mu$m.
This ensures that features consistent with dust that are not removed by the simple extinction laws employed do not influence the power-law fits.
All epochs of VLT X-Shooter and the JWST MIRI and NIRSPEC data used in our fits are shown in Figure\,\ref{fig:XS123}.

\subsubsection{X-Shooter: epoch i}\label{sec:X-shooter}
The first epoch of the X-Shooter observations is at 11$\angdot[h]$54 ($0.49$\,days) after the \fermi trigger, and has been previously used for the analysis in \citet{malesani2023, levan2023}.

\subsubsection{X-Shooter: epoch ii}
The second epoch of VLT X-Shooter spectroscopy is at $\sim4$ days post burst.
The epoch ii data was reduced at the same time and in the same manner as that of epoch i, as described in \citet{levan2023, malesani2025}.
The transient is solidly within the afterglow phase at epoch ii, where the light curve is fading and the spectrum is non-thermal.

\subsubsection{X-Shooter: epoch iii}
The third epoch of VLT X-Shooter spectroscopy was taken at $\sim10.4$\,days post burst, originally presented in \citep{malesani2025}.
At lower SNR, the third epoch exhibits much more noise.

\subsubsection{JWST: 13.3 days}\label{s:JWST_fiiting}
The first ever JWST spectra of a GRB afterglow was made by \citet{levan2023} to search for early and red emission from an energetic GRB-SN and potential signatures of $r$-process nucleosynthesis. However, these observations did not reveal any SN contribution at the time of the observations, and only a faint trace of SN emission was subsequently detected at $\sim$170 days by \citet{blanchard2024} without any additional $r$-process powered component.
The JWST data at 13 days are fit in the same way as each of the X-Shooter spectra, using the same three models and considering only the MW extinction as a ``total extinction'' value.

\begin{figure}
    \includegraphics[width=\columnwidth]{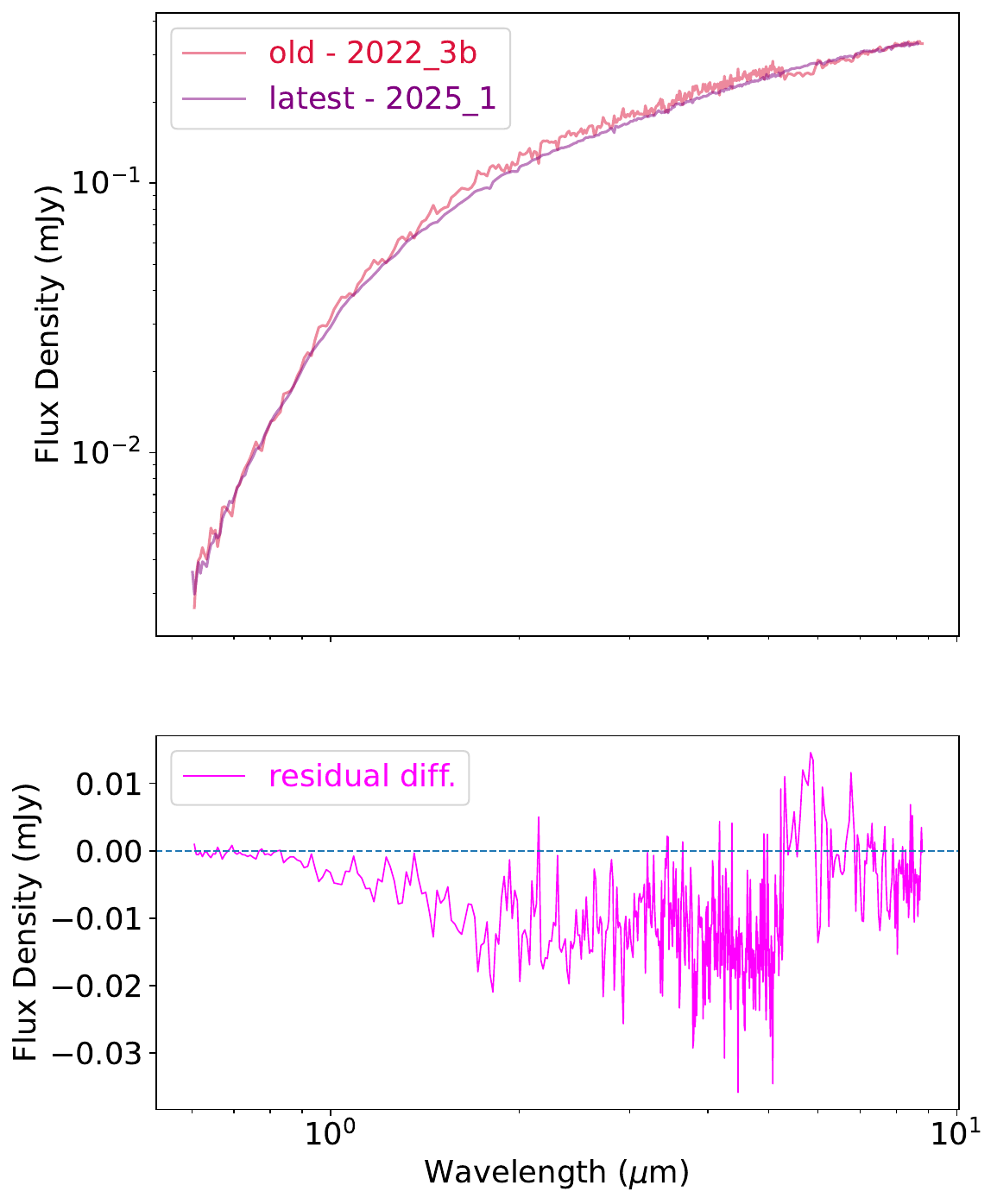}
    \caption{Difference between the old pipeline treatment of JWST data \& latest pipeline treatment of JWST data.} 
    \label{fig:jwst_datadiff}
\end{figure}

The JWST data presented in \cite{levan2023} was obtained early on in the mission lifetime and used the pipeline products available at the time of writing (in particular, the {\tt 2022$\_$3b} software version). However, subsequent updates and improvements have been made, and for this project we have re-retrieved the products from MAST which were processed with the {\tt 2025$\_$1} version of the pipeline. 
The difference between the old \citep{levan2023}, and improved data is shown in Figure~\ref{fig:jwst_datadiff}. 
The resulting difference is that the overall signal has a slightly reduced flux density when compared to the old pipeline values, although the improvements remove a previous disconnect between NIRSPEC and MIRI. 
A notable feature of the new pipeline reduction is the much improved signal to noise ratio across the entire wavelength range, resulting in a much smoother spectrum. Indeed, several plausible features in the original spectrum are now absent.  

\subsection{Model fitting}\label{sec:fit}
During the early afterglow phase, based on GRB afterglow models \citep[e.g.,][]{sari1998}, the expected spectrum is non-thermal with a spectral index $\beta$.
From a few hours to $\sim 14$ days post-burst, the BOAT's afterglow was found to be entirely non-thermal, without any evidence for supernova contribution \citep{levan2023}.
We model this with either a single power-law model (PL) $F_\nu\propto \nu^{-\beta} = (\lambda/c)^\beta$, or one of two physically motivated smoothly broken power-laws (SBPLs).
In afterglow models, there are several characteristic frequencies, hence the whole spectrum consists of several power-law segments that smoothly join at these break frequencies.
The smoothly joined power-law segments are each described with a unique sharpness parameter, $s$, that determines how rapidly the power-law changes from the short to the longer wavelength regime following the description in \citet{granot2002}.
The simple power-law model will be fitted first, followed by two different smoothly broken power-laws, as described below.

The PL is simply defined as,
\begin{equation}
    F = f\left(\frac{\lambda}{\lambda_b}\right)^{\beta}
\label{eq:simp}
\end{equation}
where $f$ is the flux constant and $\beta$ is the spectral index

The first of the SBPLs (SBPL1) consists of two power-law segments:
$\lambda^\beta$ at short wavelengths, and $\lambda^{-1/3}$ at longer wavelengths -- the spectral break in SBPL1 is equivalent to the synchrotron peak wavelength,
\begin{equation}
    F = f\left(\frac{\lambda}{\lambda_b}\right)^{\beta_1}\left[1+\left(\frac{\lambda}{\lambda_b}\right)^{s(\beta_1+1/3)}\right]^{-1/s},
    \label{eq:low}
\end{equation}
where $f$ is a flux constant, and $s = 1.84 - 0.40(2\beta+1)$.

The second of the SBPLs (SBPL2) consists of two power-law segments: $\lambda^{(\beta+1/2)}$ at short, and $\lambda^\beta$ at long wavelengths -- equivalent to the cooling break transition for our non-thermal spectrum.
The cooling break lies below the synchrotron peak and occurs towards the shorter wavelength range of the afterglow.
The functional form for this case is
\begin{equation}
    F = f\left(\frac{\lambda}{\lambda_b}\right)^{\beta_2+1/2} \left[1 + \left(\frac{\lambda}{\lambda_b}\right)^{s/2}\right]^{-1/s},
    \label{eq:high}
\end{equation}
with a sharpness parameter of $s = 1.15 - 0.06(2\beta+1)$. Both sharpness parameters are calculated by \citet{granot2002}

The PL model has the flux density, $f$, normalised to the smallest wavelength in the dataset fit as a free parameter alongside the spectral index, in the range $-2.0\leq\beta\leq2.0$ -- this allows for a broad range of potential solutions.
In each of the SBPL cases, the break wavelength, $\lambda_b$, is treated as a free-parameter that lies between $0.001< \lambda_b < 1500$\,$\mu$m. The model flux-density, $f$, is equivalent to the normalisation at the break wavelength and the spectral index is in a limited range, $0.0\leq\beta\leq1.0$ for both cases, which is then equivalent to the accelerated electron distribution index ($p$) of $1.0 \leq p \leq 3.0$. This is viable yet broader than the usually considered range \citep[often limited to $p>2$ for convenience, although examples of model fits with $p<2$ exist within the GRB population e.g.,][]{li2026}. 
For the two SBPL models, the spectral index $\beta$, always describes the $(p-1)/2$ component of the synchrotron spectrum, where $p$ is the accelerated electron distribution index, \citep[see][]{sari1998}.

\begin{table*}
    \centering
    \caption{Literature values of extinction, reddening, and the spectral index at optical and/or the electron spectral index when given for the early afterglow of GRB\,221009A. The extinction values refer to the $A_V$ extinction unless otherwise stated. Values in parenthesis indicate the inferred value from the listed $\beta$ or $p$ in the publication. Where `narrow' or `wide' are specified, this indicates the value found by the authors for a narrow or wide jet or structured jet component. Where available, the extinction values are those for the combined extinction: MW and host treated as a single value due to the MW extinction's dominance.}
    \renewcommand{\arraystretch}{1.11}
    \begin{tabular}{c|c|c|c|c|c}
    Work & $A_V$ & $E(B-V)$ & $R_V$ & $\beta$, $([p-1]/2)$ & $p$, $(2\beta+1)$ \\
    \hline
    \citet{fulton2023} & $A_r = 4.64$ & - & $3.1$ & $0.8$ & -, (2.6)\\
    \citet{kann2023} & $5.202\pm0.085$ & $1.69\pm0.03$ & $3.1$ & -, $(0.501,~0.715)$ & $2.003^{+0.005}_{-0.003}$ [narrow], $2.43^{+0.03}_{-0.02}$ [wide] \\
    \citet{laskar2023} & $4.1034$ & $1.32$ & $3.1$ & -, $(0.765)$ & $2.53\pm0.01$  \\
    \citet{levan2023} & $4.935\pm0.005$ & $1.680\pm0.003$ & $2.938\pm0.008$ & $0.362\pm0.001$, $(0.3)$ & $\sim 1.6$, $(1.724)$ \\
    \citet{malesani2025} & $4.177$ & $1.347$ & $3.1$ & $0.8$ or $0.4$ & -, $(2.6)$ \\
    \citet{shrestha2023} & $4.1$ and $5.4$ & $1.32$ and $1.74$ & $3.1$ & $0.59\pm0.17$ & -, $(2.18)$ \\
    \citet{srinivasaragavan2023} & - & $1.31^{+0.06}_{-0.07}$ & - & - & - \\
    \citet{sato2023} & $A_r = 4.31$ & - & - & -, $(0.85,~0.7)$ & $2.7$ [narrow], $2.4$ [wide] \\
    \citet{blanchard2024} & $4.63^{+0.13}_{-0.64}$ & $1.09^{+0.18}_{-0.20}$ & $4.24^{+0.74}_{-0.64}$ & $0.41\pm0.01$ & -, $(1.82)$ \\
    \citet{kong2024} & - & $1.36$ & - & - & - \\
    \citet{ren2024} & - & $1.32$ & - & -, $(0.673,~0.573)$ & $2.345\pm0.075$ [narrow], $2.145^{+0.020}_{-0.037}$ [wide] \\
    \citet{sanchez-ramirez2024} & $4.03\pm 0.19$ & $1.30\pm 0.06$ & $3.1$ & $0.579\pm0.022$, $(0.3)$ & $1.6$, $(2.158)$ \\
    \citet{sears2025} & $4.63$ & $1.09$ & $4.24$ & $0.76$ & $2.52\pm0.14$, $(2.52)$ \\
    \citet{sato2025} & - & - & - & -, $(0.6)$ & $2.2$ [narrow and wide] \\
    \end{tabular}
    \label{tab:refs}
\end{table*}

The intrinsic afterglow spectrum described by these models will undergo line-of-sight extinction and absorption at NIR/Optical/Ultraviolet (UV) and X-rays respectively.
Extinction occurs both within the host galaxy and the Milky Way due to mixtures of gas and dust between the emission site and the observer.
Similarly, absorption at higher energies, say X-ray, through the same region is due to interstellar grains' electrons interacting with oncoming high-energy photons \citep{Hoffman_Draine_2016, Costantini_2019}.
For the BOAT X-ray afterglow, the complex scattering and absorption effects have been solved in \citet{williams_maia2023}.
However, the same cannot be said for other wavelengths. An incomplete resolution of the extinction will have a noticeable effect onto model inferences of afterglow parameters. 


To avoid any bias in our analysis, we take into account all the previous literature and use our observed data to find a reasonable range for the reddening, see Section \ref{sec:fit} for details.
The aim of this analysis is to determine the intrinsic spectrum for the afterglow at optical and NIR wavelengths within the first 15 days.
Additionally, the total extinction and reddening for a fixed extinction law will be determined that can be utilised for further analysis of the afterglow at later times.

For our analysis we use the observed spectral colour to better inform the degree of reddening, $E(B-V)$.
The expected range for the spectral index between $B$ and $V$ for the BOAT's afterglow is $0.2 \leq \beta \leq 0.9$ \citep[see][etc.]{laskar2023, levan2023, shrestha2023, sanchez-ramirez2024}. 
This gives an intrinsic colour in the range, $0.04 \leq (B-V)_0 \leq 0.2$, based on the magnitudes at the effective wavelength for the bands [$B$ : 0.4448 $\mu$m, $V$ : 0.5505 $\mu$m].
Using the magnitude from the integrated $AB$ bandwidth for $B$ and $V$ bands from the epoch i X-Shooter spectrum (see Section\,\ref{sec:X-shooter}), the observed colour is $(B-V) = 1.55\pm0.63$, and is consistent across all X-Shooter epochs.
Using these colours, the $E(B-V) = (B-V) - (B-V)_0 = [1.55\pm0.63] - [0.12\pm 0.08] = 1.43 \pm 0.64$, giving a viable reddening range of $0.79 \leq E(B-V) \leq 2.07$.
We use this range for the reddening to inform our choice of prior range in the model fits.

The three power-law models, Equations \ref{eq:simp}, \ref{eq:low} \& \ref{eq:high}, are fit to each epoch of the X-Shooter and JWST spectra using {\sc Nessai} \citep{williams2021, williams2023}, a nested sampling algorithm for Bayesian inference that incorporates normalising flows, via {\sc Bilby} \citep{bilby_paper}.
As noted in Table\,\ref{tab:refs}, the results of model fits to the data give varied results depending on the assumptions made in each case.
The exact values for the total extinction, $A_V$, the reddening, $E(B-V)$, and the spectral index at optical/NIR, $\beta$, are particularly poorly constrained.
To examine this, we fit each model to the data from each epoch with a fixed extinction.
We repeat these fits for a new fixed extinction value $\Delta A_V = 0.5$ greater than the previous until we have covered the range $3.5\leq A_V\leq 7.0$ for each model and epoch.
For each fit we marginalise over our fiducial reddening range while ensuring that the ratio remains between $2.0\leq R_V \leq 6.0$ as a condition.

Due to the highly extinct line-of-sight, the curvature of the extinction law and a PL model can become degenerate, especially at the optical and bluer wavelength ranges.
Where our fits show a steep linear trend between $\beta$ and $A_V$, this is indicative of a spectrum vs extinction degeneracy, and should be clearest in the PL model.

Due to the low inferred host extinction (when compared to the MW line-of-sight) \citep[e.g.,][]{levan2023, laskar2023, oconnor2023, ren2024, sanchez-ramirez2024}, the associated uncertainty on the exact host extinction law \citep{kann2023}, and the low-redshift of the host \citep{malesani2025}, the choice of host extinction is shown by \citet{srinivasaragavan2023} to have little effect on model fit parameters.
We therefore use a combined `total' extinction throughout, assuming a \citet{fitzpatrick1999} law which allows us to vary $R_V$ (and therefore $E(B-V)$ reddening) for each model fit for fixed $A_V$.
Given the wavelength range of the data, the choice of extinction law will have minimal impact on the resulting values.

The results of these fits are listed in Table\,\ref{tab:XSe1_fit}. 
The log evidence values for each model are compared in Figure\,\ref{fig:log_e} (top panels).
Where the evidence\footnote{The log evidence values can be both positive and negative, and can only be used to infer model selection for models tested against the same dataset.} is highest (i.e., the y-axis value goes to 1), a refined, higher resolution fit to data is performed with $\Delta A_V = 0.05$, and the combined results are plotted in the bottom panel of Figure\,\ref{fig:log_e}. 

\subsubsection{An ``Holistic'' Approach: Combined fitting}\label{sec:combined_fit}
Following the initial individual epoch and three model (PL, SBPL1, and SBPL2) fits to the early spectral data, we utilise a more complex triple power law spectral model and simultaneous fitting to all epochs.
The single epoch model fits reveal qualitative trends across the four epochs and highlight the degeneracy in the extinction-spectral-index parameters.
The individual epoch-model fit analysis revealed:
Extinction in the range $3.8\leq A_V \leq 5.05$, a mean extinction from the X-Shooter epochs (i), (ii), and (iii) of $A_V\simeq 4.6$, and a JWST average of $A_V\simeq4.1$ giving an all epochs mean of $A_V\simeq4.46^{+0.59}_{-0.66}$ and consistent with the range of extinction from community analysis (see Table\,\ref{tab:refs}).
The model fits to the X-Shooter epochs and the JWST epoch all prefer an SBPL model that follows a GRB afterglow synchrotron peak frequency broken power-law, e.g., Equation \ref{eq:low} (SBPL1) and a fixed extinction $A_V=4.0$ (see Tables\,\ref{tab:bayes}
 \& \ref{tab:other_bayes}, where model preference is found via the log Bayes factor comparison).

A more ``holistic'' approach to fitting all the data used within this study is to perform the sampling simultaneously, allowing only a single extinction, spectral index, and a spectral-density model that includes all the elements of our original three models.
In fitting the full spectral data set, we allow the extinction to be a free parameter in the broader range $2.5\leq A_V\leq7.0$, and the reddening in the range $0.7\leq E(B-V)\leq 2.2$, while ensuring that the ratio remains between $2.0\leq R_V\leq 6.0$, and we fit a double-smoothly-broken power-law (DSBPL) model.
This DSBPL has a single value for the spectral index of $0.0<\beta<1.0$ between two breaks and uses a shared $A_V$ and $E(B-V)$ value for all epochs, while the two break frequencies are defined as varying with time as, $\nu_m\propto t^{-3/2}$ and $\nu_c \propto t^{-1/2}$ or $t^{1/2}$ for either a uniform or wind density profile.
The smoothness of the breaks are defined by those listed in \citet{granot2002} for either a uniform or wind medium: uniform has $s_1 = 1.84 - 0.4 p$, and $s_2 = 1.15 - 0.05 p$; and wind has $s_1 = 1.76 - 0.38 p$, and $s_2 = 0.8 - 0.03 p$.
We ensure that $\nu_m<\nu_c$, and the spectrum is slow-cooling throughout the epochs considered.

The DSBPL is defined as,
\begin{equation}
    F_\nu = C \left[\left(\frac{\lambda}{\lambda_m}\right)^{- s_1 /3} + \left(\frac{\lambda}{\lambda_m}\right)^{-\beta s_1} \left[1 + \left(\frac{\lambda}{\lambda_c}\right)^{-s_2 /2}\right]^{s_1/s_2}\right]^{-1/s_1}
    \label{eq:DSBPL}
\end{equation}
where we have shown the equation as a function of wavelength, $\lambda$, instead of frequency for consistency with the earlier expressions\footnote{This is also the format for how the function is fit e.g., for flux density, $F_\nu$, and wavelength, $\lambda$. The conversion to frequency space is trivially, $\nu = c/\lambda$.}.
The subscript $m$ and $c$ indicate the synchrotron and the cooling breaks respectively, while $s_1$ and $s_2$ are the smoothness parameter in each case.
The formalism of the DSBPL is that for slow cooling with $F_\nu(\nu<\nu_m) \propto \nu^{1/3}$, $F_\nu(\nu_m<\nu<\nu_c) \propto \nu^{-\beta}$, and $F_\nu(\nu>\nu_c) \propto \nu^{-\beta-1/2}$ with $\beta = (p-1)/2$. The step-by-step derivation of this model is done in Appendix \ref{appd:C}.

\section{Results}\label{sec:results}
\begin{figure*}
    \centering
    \includegraphics[width=\textwidth]{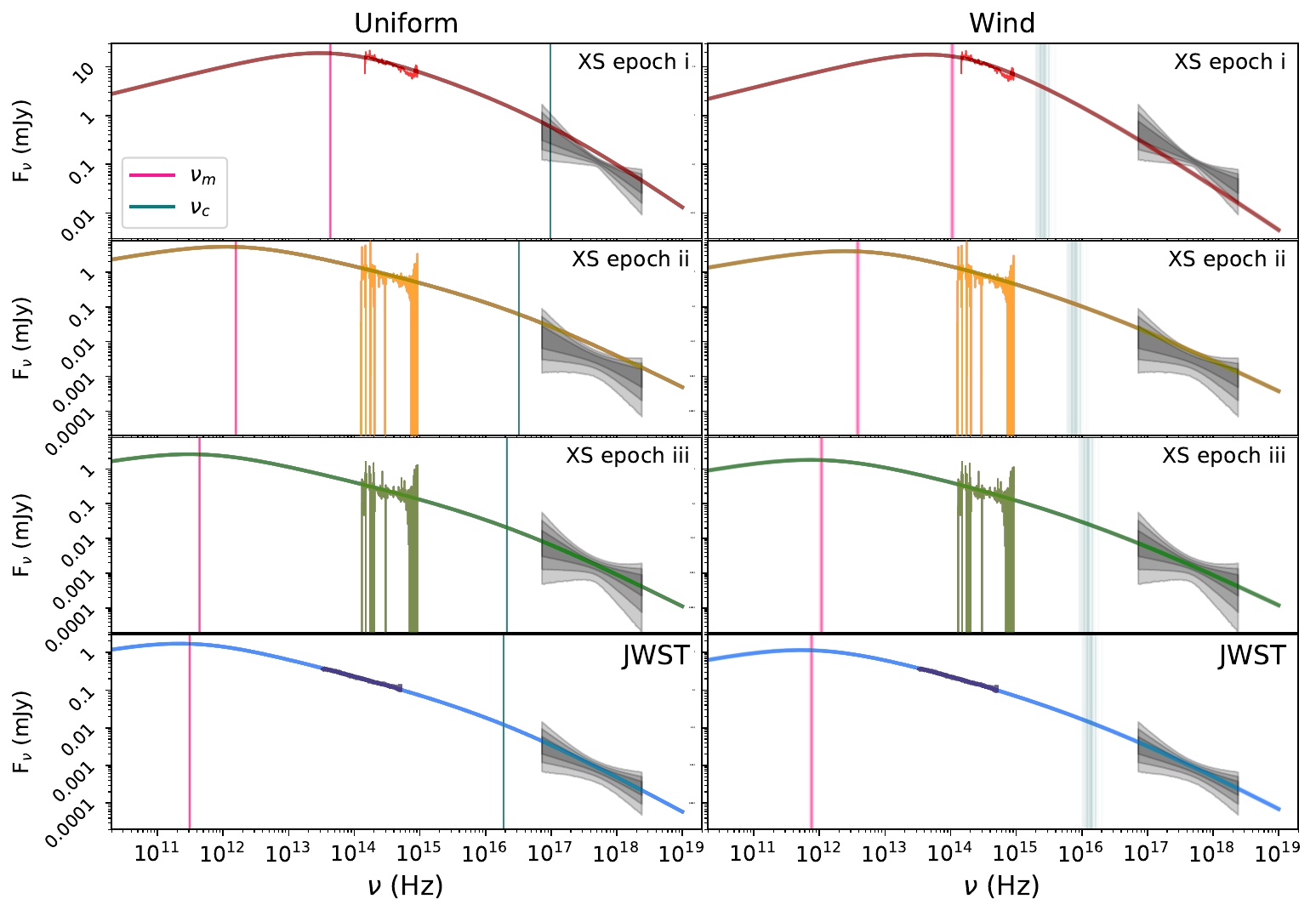}
    \caption{The joint, double-broken power-law fit to all spectral data considered. Each panel shows 400 randomised model draws and the extinction-removed spectral data profiles. The  axis is extended beyond the range of the fits to show the inferred X-ray spectral slope. The grey butterflies indicate the XRT SEDs at the nearest time to the individual epochs. \textbf{Left}: The panels show the results for a uniform interstellar medium, using the appropriate smoothness parameters for the uniform case \citep[from][]{granot2002} and a cooling break that evolves with time as $\nu_c \propto t^{-1/2}$. The spectral index for $\nu_m<\nu<\nu_c$ is $\beta=0.45$, with a total extinction, $A_V=4.4$ and a reddening of $E(B-V) = 1.31$, while a broad prior was used for the possible $\nu_m$ (pink vertical line), we constrained $\nu_c$ (teal vertical line) to be less than 1 keV in the first epoch -- motivated by the highest evidence fits to individual epochs using a single of smoothly broken power-law model. This ensures consistency with the observed X-ray spectral index, $\beta_X\sim 0.9$ (note that leaving $\nu_c$ unconstrained resulted in a best fit value for the cooling break in the GeV energy range and an inferred $\beta_X\sim0.44$, inconsistent with observations during the time frame considered here). \textbf{Right}: The panels show the results for a wind medium defined as $n(r) \propto r^{-2}$. Here the cooling frequency evolves as $\nu_c\propto t^{1/2}$ and a restricted range on the $\nu_c$ prior is not required. Spectral index in $\nu_m<\nu<\nu_c$, the total extinction, and reddening are $\beta=0.42$, $A_V=4.33\pm0.01$, and $E(B-V)=1.28$ and are consistent with, although marginally smaller, than the values for the uniform medium case.}
    \label{fig:combo}
\end{figure*}

We fit the DSBPL model using \texttt{Nessai}, as for the individual epoch fits, using the prior ranges for parameters listed in Table \ref{tab:func4}.
A random sample of the spectral model fits verses the data are shown in Figure\,\ref{fig:combo} -- epoch (i) has a red line, epoch (ii) a yellow or gold line, epoch (iii) a green line, and the JWST epoch a blue line.
The data at each epoch is shown without the associated uncertainty, however, this was used in the joint model fitting.
The position of the cooling, and the synchrotron breaks are shown with teal and pink vertical lines, respectively.
The change in break frequencies with time for the two environments is clearly shown; where the synchrotron frequency has the same behaviour regardless of medium and the cooling frequency has a medium dependence that results in an opposite proportionality with time.

The model fits return a total extinction and reddening of $A_V = 4.40 (4.33)$ and $E(B-V)= 1.31(1.28)$ for a uniform(wind) medium\footnote{The uniform medium result is consistent with the \citet{schlafly2011} reddening, $E(B-V) = 1.32$ for the sky localisation -- see Section\,\ref{sec:methods}.} with the inferred ratio, $R_V = 3.37(3.39)$.
For the BOAT, MW extinction dominates the extinction \citep{srinivasaragavan2023}, and some studies find a negligibly small host galaxy extinction \citep[e.g.,][$A_{V,{\rm host}} = 0.02$]{levan2023, malesani2025}.
However, given the \citet{schlafly2011} table value for the sky location, $A_V/E(B-V) = 2.74$, and assuming that the reddening is dominated by the MW, with $E(B-V) = 1.32$, then the line-of-sight extinction is $A_{V,{\rm~ MW}} \simeq 3.58(3.50)$, and the difference in the total extinction and MW is that attributed to the host.
The host galaxy optical extinction found  by other authors via optical photometry and SED analysis is varied, however; \citet{ren2024} finds $A_{V,{\rm ~host}} = 0.88$, \citet{blanchard2024} finds $A_{V,{\rm ~host}} =0.67^{+0.11}_{-0.07}$, and \citet{kann2023} finds $A_{V,{\rm ~host}} = 0.71\pm0.07$, while our results infer a host extinction of $A_{V,{\rm ~host}} \simeq 0.8(1.0)$ uniform(wind), which is consistent with these previous estimates showing small, but not insignificant, host extinction.

In Figure\,\ref{fig:combo}, we extend the frequency scale to include the 0.3 - 10 keV range observed via {\it Swift}-X-Ray Telescope (XRT).
The data is the unabsorbed, X-ray flux from the UK {\it Swift} Science Data Centre\footnote{\href{https://www.swift.ac.uk/xrt_spectra/01126853/}{UK Swift Science Data Centre, University of Leicester, Leicester, UK.}}.
As no spectral break is seen to pass through the X-ray bands in these four epochs, we fix the spectral index as that given by the {\it Swift} Science Data Centre, a photon index of $\Gamma = 1.73\pm0.17$ or spectral index, $\beta = 0.73\pm0.17$.
The grey shaded butterflies indicate the 1-, 2- and 3-$\sigma$ uncertainty in the flux density across the spectrum.
Note that the DSBPL is not fit to the XRT data, and the coincidence of the extrapolated spectra can be taken as evidence for the strength of the optical and NIR spectral fits i.e., they give a reasonable prediction for the observed X-ray spectra at the time of these observations.

\begin{table}
    \centering
    \caption{The prior and posterior ranges for the extinction, reddening, spectral index and break wavelengths for the uniform and wind medium fits to all the optical and NIR spectral data within the 11 hours and 13 days post burst for GRB 221009A with the DSBPL model, Equation \ref{eq:DSBPL}. The prior distribution is linear between the values listed in parenthesis i.e., $\lambda_{c,m}$ are both log uniform. The break wavelengths, $\lambda_{c,m}$, are defined as the model values at the JWST epoch or 13.2 days post burst. The `information' score, $\Delta_i$, comparison between these two fits marginally favours the wind medium model over the uniform medium (where $2<\Delta_i<4$ indicates that there is considerable support for the model).}
    \label{tab:func4}
    \begin{tabular}{c|c|c|c}
        Parameter    & Prior &  Uniform & Wind \\
        \hline
        $\beta = (p-1)/2$  & $(0.0,~1.0)$  &  $\bf 0.447\pm{0.001}$ & $\bf 0.425\pm{0.003}$  \\
        $\lambda_c$ ($\mu$m) & $10^{(-6.0, ~6.0)}$ & $0.016\pm0.000$ & $0.022\pm0.003$ \\
        $\lambda_m$ ($\mu$m) & $10^{(-6.0,~6.0)}$ & $968\pm13$ & $395\pm13$ \\
        $A_V$ (mag) & $(2.5,~7.0)$ & $\bf 4.402\pm {0.004}$ & $\bf 4.326 \pm {0.007}$\\
        $E(B-V)$ & $(0.7,~2.2)$ & $1.308\pm0.004$ &  $1.276\pm0.005$ \\
        \hline
        $\Delta_i$ & -- & $3.33$ & $0.0$
    \end{tabular}

\end{table}

\begin{figure}
    \centering
    \includegraphics[width=\columnwidth]{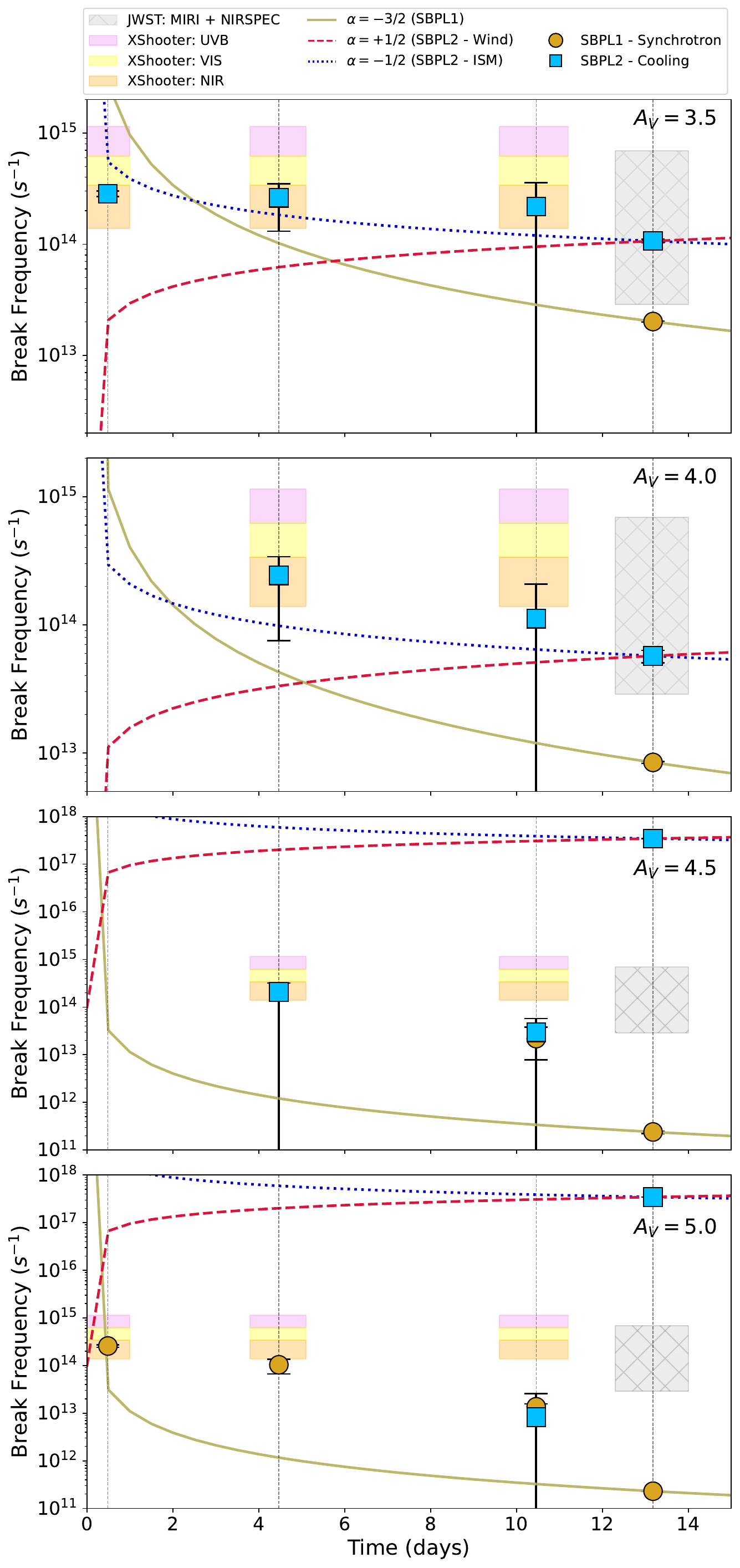}
    \caption{The break frequency evolutions of the 0.5, 4.0, and 10.5 day X-Shooter and the 12 day JWST spectra. These specific values represent the potentially viable options that were obtained via Figure~\ref{fig:XS123}, with the extinctions of $A_v = 3.5, 4.0, 4.5,$ \& $5.0$ split into four panels arranged vertically in their respective order. The frequency coverage of each instruments are layered in the background to each epoch to demonstrate whether the data is within the means of observations. Their break frequencies are obtained via fittings of two different smoothly-broken power law (SBPL) models, a synchrotron and a cooling break model. Their selections are made based on whether their electron index, $\beta$, lies within the expected physical range of $0.20 \lesssim \beta \lesssim 0.90$. Their time evolutions are extrapolated using the 12-day JWST fitted spectra as the anchor point, and are based on their associated SBPL models, with $\nu_b \propto t^{\alpha}$, and $\alpha = -3/2; -1/2; +1/2$, corresponding to SBPL1, SBPL2 (uniform ISM) \& SBPL2 (wind-rich), respectively.}
    \label{fig:break_freq_evo}
\end{figure}

\section{Discussion}\label{sec:discussion}
We have re-analysed the optical and NIR spectra of the afterglow to the BOAT within the first 14 days -- this timescale is critical to understand the afterglow behaviour as it covers the brightest afterglow emission.
Due to an extremely dusty sight-line, afterglow analysis has resulted in multiple inconsistent results (see Table\,\ref{tab:refs}).
However, via the spectral analysis at four epochs, we have found sets of potentially viable solutions, and recreated the range of extinction and reddening values found by the community, $3.80 \leq A_V \leq 5.05$ and $0.79 \leq E(B-V)\leq 2.09$ respectively.

In Figure~\ref{fig:AV_xb} we show the relationship between the afterglow spectral index ($\beta$) and the visual extinction ($A_V$).
The top panel shows the results for the simple power-law model, while the middle and bottom panels show the results of SBPL1 and SBPL2 model fits respectively. 
The model SBPL1 is motivated by the synchrotron break, $\nu_b$ (middle), and the SBPL2 model represents the cooling break, $\nu_c$ (bottom). 
The top panel, showing results from the simple power-law fits, clearly demonstrates a strong correlation between extinction and the inferred spectral index across all the optical X-Shooter epochs, $\beta \propto A_V$ or the inferred spectral index is a function of the assumed extinction.
At optical and bluer wavelengths, the effects of extinction are increased, and where the line-of-sight extinction is large (as is the case for the BOAT) then the curvature of the extinction law and the spectral index become degenerate -- this is less of an issue at the NIR of the JWST NIRSpec and MIRI spectra.

For these individual model fits to the data, we can perform model selection on the sample.
Using the JWST epoch fits, the log Bayes factors are shown in Table\,\ref{tab:bayes}, with the best model for each row and column shown with boldface text.
The JWST epoch data is best fit at an extinction of $A_V  = 4.0$ by the SBPL1 model.
Further, by fixing the extinction to this value, we show the model selection for the three X-Shooter epochs in Table\,\ref{tab:other_bayes}; epochs i, ii, and iii all prefer the broken power-law model SBPL1 over either PL or SBPL2.
As the X-Shooter spectra cover a bluer wavelength range, for $\nu > \nu_m$ as required by the JWST epoch model, and as $\nu_m \propto t^{-3/2}$ (moving redder with time), then the single power-law preference at epochs i, ii, and iii requires $\nu_b < \nu_{\rm red}$.
At the JWST epoch, the best fit SBPL1 spectral break is at $\lambda_b = 40.87~\mu$m or $\nu_b = 7.3\times 10^{12}$\,Hz, given the timescale, this break frequency would be at $\nu_{b,i} \sim 10^{15}$\,Hz, and in the UVB range of the X-Shooter arm -- we do not see this for the $A_V = 4.0$ and epoch i best fit data.

To visualise how each of these SBPL models evolve, we select values of break frequencies that have fitted spectral indices within the physical range at each epoch. 
We plot these against time post-burst in Figure~\ref{fig:break_freq_evo}. 
At each epoch, we indicate the observational limits of the instruments used, to illustrate whether the recovered break frequency lies within, or near, the observable range. 
We label the break frequencies associated with a synchrotron break, fitted with SBPL1, with a yellow circle; and those of the cooling break with a blue square. 
We then extrapolate three different scenarios of the SBPL models using the 13-day JWST spectra as the anchor point. This is chosen simply because of its superior SNR compared to the X-Shooter epochs. 
For the cooling break model, we consider two environmental scenarios; one in which the jet propagates through a uniform density medium (ISM), and another where the density decreases as a function of the radius (wind-like medium) \citep{chevalier1999}. The break frequencies are then evolved in each scenario using:

\begin{equation}
    \nu = \left( \frac{t}{t_{\text{epoch}}}  \right)^{\alpha} \nu_{\text{epoch}},
    \label{eq:break_freq_evo}
\end{equation}
where $t$ is the time at which the frequency ($\nu$) is being calculated, $\alpha$ is the power index that varies depending on the SBPL scenario and follows $\alpha = [-3/2, -1/2, 1/2]$ for SBPL1, SBPL2 (ISM) and SBPL2 (Wind), respectively \citep{granot2002}.

From $A_V = 3.5-4.5$, the cooling breaks (blue squares) show a break that trends to lower frequencies with time -- this decline is consistent with an afterglow propagating into a uniform ISM.
For $A_V = 5.0$, there are insufficient cooling breaks to determine a clear trend.
Where a viable synchrotron break (yellow circle) is plotted for sufficient epochs, the trend is similarly declining, as expected.

At $A_V = 4.5-5.0$, there is a sharp spike in the cooling break frequency and a sudden drop-off for the synchrotron peak frequency, at the JWST epoch.
These values for the break frequencies are at the limits of the prior range and indicate that at this time, the data are strongly within the power-law segment of the SBPL1 and SBPL2 models -- where a low and a high break frequency leave the data to be fit by $\beta = (p-1)/2$ component of the model.
None of the break frequency series follow a perfectly self-consistent evolution between the epochs, as each are found independently of the other epochs.
Utilising the JWST epoch as an anchor, the trend across all epochs is approximate, however, it is still qualitatively useful in arguing against a wind medium -- where the cooling break would be seen to follow an increasing trend with time.

\subsection{Light curve expectations versus reality}

Given a spectral index, an assumption of the spectral regime, and the surrounding medium density profile, the standard afterglow closure relations can be used to infer the afterglow flux density temporal index \citep{gao2013}.
In modelling the afterglow light curve, the observed decline indices lend themselves to be interpreted as a slow cooling, wind medium afterglow with electron index $p\sim2.5$ \citep[e.g.,][]{fulton2023, laskar2023, sato2023, ren2024, sears2025}. 
However, such an interpretation is never consistent with the spectral index found from the early X-Shooter and JWST observations \citep[see,][and this work]{levan2023, malesani2025}.

Late spectra appear to have redder colour than these early observations \citep[e.g., the spectral index at $\gtrsim 50$ days is consistent with $\beta\sim0.8$, as in][although this is not definitively measured but inferred from the mm to X-ray extrapolation]{blanchard2024}; 
this would be the expectation for the passage of the cooling break in a uniform medium, as the shallow to steeper spectral energy distribution is consistent {\it only} with this scenario in standard GRB afterglow physics.

If we consider the uniform medium, where the cooling frequency moves to lower frequencies with time as $\nu_c\propto t^{-1/2}$, then the expected position of the cooling break at $ 50\pm 10$\,days post burst will be $\sim9.17^{+1.2}_{-0.8} \times10^{15}$\,Hz; this places the cooling break from this model analysis in the UVB range at $\sim50$\,days post burst.
A late jet break was claimed by \citet{sears2025} based on later optical data (at approximately half a year to one year post burst), the temporal position of a jet break was estimated to be $\sim 50\pm10$\,days, although no corresponding break is seen at lower, radio frequencies \citep[out to very late times][]{rhodes2024}. 
We therefore suggest that the apparent break identified by \citet{sears2025} is not the jet break but the passage of the cooling break through the optical regime.
In fact, if we allow our model to have a sharp cooling break, then a break that passes into the optical at $\sim50$\,days is easily achieved -- it is the smoothness of this cooling frequency that requires a ``higher'' initial value (e.g., see the wind medium model which is best fit with a cooling break at blue/UV frequencies in epoch (i), implying that a sharper cooling break can easily be accommodated by the data).
As the temporal break information at $\sim 50$\,days is not included in our model fit, any recreation of this scenario by our models is coincidental (or predictive), however, if this break is indeed the cooling break, then the standard wind medium is instantly ruled out.

The analysis in Section 5.1 of \citet{levan2023} still qualitatively holds i.e., the closure relations can reproduce the observed temporal decline rates for the data from $\sim$0.5 to $\sim15$ days at near-infrared, optical and X-ray only for the case of a uniform medium, post-geometric\footnote{Geometric refers to the jet break case where the steepening of the afterglow decline is due only to the edge effect without any lateral or sideways expansion.}-jet-break.
For this scenario, a uniform medium and $\beta=0.447$, for $\nu_m<\nu<\nu_c$ has a shocked electron distribution index of $p = 1.894 < 2$.
The analytic decline indices for emission at $\nu<\nu_c$ is $\alpha_{\nu<\nu_c} = 3(p + 6) / 16 = 1.48$, and for $\nu>\nu_c$ has $\alpha_{\nu>\nu_c} = (3p + 22) / 16 = 1.73$, and both consistent with the data \citep[e.g.,][]{shrestha2023, williams_maia2023}.

\begin{figure}
    \centering
    \includegraphics[width=\columnwidth]{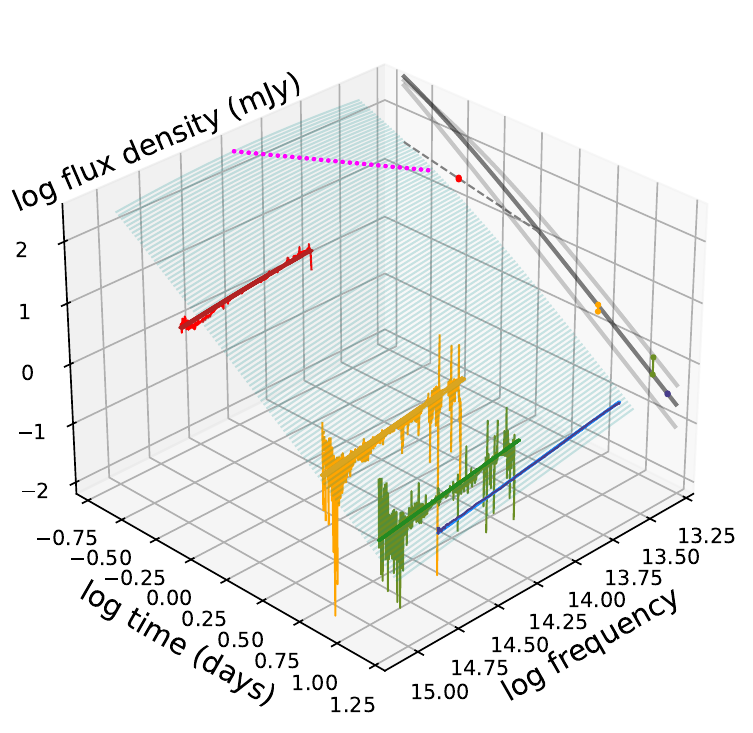}
    \caption{Three dimensional frequency (x-axis), time (y-axis) and flux density (z-axis) plot that shows the 4 epochs (red -- 0.5 days, orange - 4.5 days, green - 10.5 days, and blue - 13.2 days) of spectroscopy used in this work. The model fit spectra using the double-smoothly-broken power-law (DSBPL), and the inferred temporal and spectral evolution shown as a dark blue-grey surface. The location of the synchrotron peak frequency is shown with magenta dots, and the light curve at $\sim 1.6\times 10^{14}$\,Hz is shown on the y-z surface as a dark grey line, with the limits of the surface, at $3.2\times10^{13}$ and $10^{15}$ Hz shown for reference. The coloured points on the y-z plain indicate the projection of the data at $\sim 1.6\times10^{14}$ Hz. The black dashed line indicates a pre-jet-break evolution that would include the 0.5 day flux density. A break at about $\sim 0.6$ days was identified at optical \citep[see e.g.,][]{shrestha2023} and X-ray at $\sim0.9$ days \citep{williams_maia2023}.}
    \label{fig:3D}
\end{figure}

\subsubsection{A solution ?}
The NIR, optical, and X-ray decline following the BOAT can only be explained via the deceleration of a narrow (early required jet break) jet within a uniform ISM-like environment.
The total MW and host extinction dominates the optical emission with an $A_V = 4.402\pm0.004$ and an $R_V = 3.365\pm0.003$ (from a fit $E(B-V) = 1.308\pm0.004$.
The spectral index at optical, where $\nu_m<\nu<\nu_c$ is best fit by a value $\beta = 0.447\pm0.001$, giving an accelerated electron distribution index, $p = 1.894\pm0.002$. 
In such a case, the high energy emission at or beyond X-ray energies may hold information about the maximum accelerated electron energy within the shock system.
As noted above, using this value for $p$, the closure relations give a decline index of $\alpha = 3(p+2)/16 = 0.73$ (where $F_\nu \propto t^{-\alpha}$), which is a much shallower decline than that observed. 
However, a geometric (edge-effect) jet break would increase this decline index by $3/4$, so that $\alpha = 3(p+2)/16 + 3/4 = 1.48$, which matches the observed decline within the data.

In Figure \ref{fig:3D} we show each of the observed epochs in frequency-time-flux density space, we additionally show a single model drawn from the posterior distribution.
Each epoch is coloured is in previous figures: red -- epoch i; yellow -- epoch ii; green -- epoch iii; and blue -- JWST epoch.
The inferred temporal evolution is shown with a grey-blue surface where the spectral shape matches that of the fit DSBPL at each time step and the amplitude of the flux density follows a $F_\nu \propto t^{-3(p+6)/16}$ evolution.
A reference frequency of $1.6\times10^{14}$\,Hz is used to demonstrate the light curve evolution. This frequency is chosen as it remains within a single spectral regime at each epoch where the light curve is projected onto the time-flux density plane with a dark grey/black line, and the flux density of the data at the reference frequency and each epoch time are shown as dots with the respective colour.
The spectral limits of the plotted surface are similarly shown with light grey lines.
Additionally, the passage of the synchrotron peak frequency, $\nu_m$ is shown with magenta coloured markers on the surface.
From inspection of this figure it can be seen that the epoch i flux density falls below the inferred temporal line, remembering that the flux temporal evolution was not a fit parameter in our model.
However, the analysis of the optical by \citet{kann2023} and \citet{shrestha2023} note that there is a break at $\sim0.6$ days, and after the epoch i observations.
They state that this is consistent with a jet break, being additionally seen at X-ray frequencies \citep[][noting $t_{X, ~{\rm break}} = 0.9\pm0.1$\,days]{williams_maia2023}.

The temporal evolution before and after this break of the optical data is consistent with that inferred via our temporal analysis;
with a model pre-break decline $\alpha_{t<{\rm break}} = 0.73$, while the data has $\alpha_{t<{\rm break}} = 0.81$, and a model post-break evolution $\alpha_{t>{\rm break}} = 1.48$ vs data with $\alpha_{t>{\rm break}} = 1.46$.
Using the epoch i reference frequency data, we show the temporal evolution of the pre-break light curve with a dashed grey/black line, the intercept of the dashed, pre-, and solid, post-jet-break light curve is seen $\sim 1.0$ days.

Similarly, the post-break temporal decline at X-ray is reproduced by the same assumptions but for an observation frequency above the synchrotron cooling frequency, $\nu>\nu_c$.
Such a scenario would naturally lead to a break at optical frequencies due to the passage of the cooling break\footnote{Remembering that the invoked jet break is the edge effect only, and so the decline index is sensitive to the cooling break.} at later times. 
Such a break is inferred by \citet{sears2025} from the very late time photometry, although they claim it is a jet-break, the optical only break has more in common with the cooling break e.g., a change in temporal index $\Delta\alpha\sim 1/4$ and a change in spectral index $\Delta\beta\sim 1/2$.
The changes in the temporal decline are lower than those inferred by \citet{sears2025}, however, we note that there is a large gap in the data from $\sim50$ days until $\sim200$ days. The host galaxy subtraction is tricky, and changes in the predicted temporal index depend heavily on the assumptions applied to modelling a complex afterglow.
A steeper temporal decline than that invoked via the combination of the edge-effect and the cooling break passage can be achieved by the onset of lateral, or sideways expansion.
In such a case, the decline index would analytically approach $t^{-p}$, where $p = 1.894$, which gives an $\alpha \sim 1.9$ and less than the $2.34$ found by \citet{sears2025}. However, numerical modelling of GRB jets demonstrates that the $F_\nu \propto t^{-p}$ underestimates the potential decline index just after the jet-break (or on-set of sideways expansion) \citep{vaneerten2013, lamb2021}. 
Such a steeper decline than the limiting case of the edge-effect is expected once sideways expansion commences. 
Invoking a jet break without any corresponding achromatic observations, either at X-ray or radio (where no break is seen in either at these times) suggests the change in decline at optical is a chromatic effect such as the passage of the cooling break inferred via the analysis in this work.

The radio data for the afterglow of the BOAT has a decline index and a spectral index that is at odds with all NIR, optical and X-ray observations and analysis \citep{laskar2023, oconnor2023}.
The radio data has been described with multiple components including reverse shocks and a long lasting forward shock \citep{rhodes2024} -- without any evidence of a jet break.
In explaining many of the complex observations related to the afterglow emission from the BOAT, many authors invoke a two-component jet \citep[e.g.,][]{sato2023, sato2025, zhang2024, zheng2024, abe2025}.

The radio and optical to X-ray discrepancy is additionally evidence for a two-component structured jet, such that the core is a narrow spine that is contained by a wide second component.
The narrow core is highly energetic and dominates the NIR/optical and the X-ray emission, while the radio emission is dominated by the wider second component structure. 
Such a scenario requires the two components to have independent microphysical parameters, so that the low frequency emission is dominated by the higher latitudes (Grabham et al. in prep), the electron index within the shock from the wider second component will have $p>2$, ensuring that the optical and X-ray emission from the second component have a lower flux density than the optical and X-ray of the core\footnote{While the opposite is true for the radio emission. Some small contribution from the wide or narrow component may add to the total flux observed at either radio or optical/NIR, and may add an additional complication.}.
This two-component structure can naturally explain many of the irregularities seen in the BOAT; from the early high energy requirements, to the radio-optical afterglow mismatch.
It additionally may hold clues to the BOAT's apparent rarity, where the chance alignment of a very narrow and energetic spine feature may produce a separate population of GRBs in the $\log N - \log S$ plane (fluence distribution), as currently understood and observed from past GRB catalogues \citep{Paciesas_1999, Paciesas_2012}. 
Thus, a more detailed and statistical investigation is required to answer questions regarding the expected rate and luminosity function of such a structured spindle/spine and broad sheath type structure.

\subsubsection{Indication of particle acceleration}
Classically, GRB afterglows are assumed to have a distribution of shocked electrons that are best described by a single power-law, $\mathrm{d}n_e/\mathrm{d}\gamma_e \propto \gamma_e^{-p}$, with $p\geq 2$ -- and more accurately, $p = 2.23$ \citep[][etc.]{achterberg2001, keshet2005} -- here $n_e$ is the number of electrons, and $\gamma_e$ is the electron Lorentz factor.
If electron acceleration in relativistic collisionless shocks of GRB afterglows is dominated entirely by the Fermi process, and the upstream magnetic field is parallel to the shock normal, then $p>2$ should hold for all GRBs.
However, observationally, GRB afterglows often have behaviour that is better described by closure  relations that have $p<2$ \citep[see][for a detailed list of the closure relations for $1<p<2$]{gao2013}.

For the BOAT we find $p = 1.894~\pm~0.002$, a value below the usually assumed range for GRB afterglows. However, as noted above, only the X-ray, optical and NIR afterglow are consistent with this low $p$ value \citep{laskar2023, levan2023} and the radio data can be fit assuming the usual $p>2$ \citep{laskar2023, rhodes2024}.
Interestingly, the photon spectral index found by \citet{klinger2024} was $\Gamma>2.2$ at energies greater than several keV, and $\Gamma\sim1.5 - 2.0$ at energies below a few keV. The low keV photon index is consistent with that found by our inference, above the cooling break, $2p + 1 = \Gamma = 1.894$, suggesting the higher energy photons may share their origin with the radio emission.
Additionally, the X-ray, optical and NIR afterglow within the first 50 days, requires a very narrow jet\footnote{A narrow jet is consistent with analysis of the very high energy data in the first minutes of the afterglow emission from the BOAT \citep{LHAASO2023, zhang2024, zheng2024, zheng2024_chao}, however, some uncertainty about the narrow jet/core interpretation for the early high energy emission has been made \citep[claiming data consistent with and without a jet break,][]{zhang2023_hai-ming} and claiming no early break required \citep{foffano2024}.}.
GRB afterglow relations require a narrow jet core to explain the temporal evolution, where a geometric loss of flux due to a jet-edge can account for the observed decline at optical and X-ray frequencies \citep[e.g.,][]{levan2023}.

Such a narrow jet core region may not follow the basic assumptions required for a Fermi particle acceleration process to result in $p>2$ e.g., the magnetic field may have a high obliquity angle, breaking the parallel field--normal assumption.
For relativistic collisionless shocks, \citet{sironi2011} found that for high-obliquity angles the electron distribution will appear shallower if the accelerated electron distribution is modelled via a single power law.
Where a magnetic-field--shock-normal is perpendicular, the electron distribution index goes to $p=1.5$.

The combination of a narrow jet core region and the required $p<2$ for the afterglow emission can provide information about the magnetic field structure, particle acceleration processes and the jet structure in collapsar origin GRBs.
These results highlight the importance of intensive photometric and spectroscopic follow-up observations of GRB afterglows.  
The JWST spectra, originally presented in \citet{levan2023} and analysed in further detail here, challenge the standard assumptions made in GRB afterglow analysis.
Additionally, these observations motivate the need for further, detailed simulations investigating shock microphysics, particle acceleration, and jet structure in GRB afterglows.

\section{Conclusions}\label{sec:conclusions}
The large line-of-sight extinction for the afterglow to the BOAT, GRB\,221009A, has resulted in a diverse spread of optical extinction, reddening, and $R_V$ values being used or assumed by the community in the analysis and interpretation of the optical and NIR afterglow data.
Using optical and NIR spectra from the first 14 days post-burst, we have shown how assumptions made about the optical spectral index and/or the extinction have resulted in a wide range of values within the literature.
Using a physically motivated double smoothly broken power-law spectral model and fitting four epochs of spectral data simultaneously, we find:
\begin{itemize}
    \item[\textbf{(i)}] If we assume that the cooling break evolves as in a uniform medium, then we find that the line-of-sight total extinction is $A_V = 4.402 \pm 0.004$. In the case of a wind medium where the cooling break increases with time, then the extinction is $A_V = 4.326 \pm 0.007$.
    
    \item [\textbf{(ii)}] In a uniform medium, the reddening via model fitting is $E(B-V)=1.308\pm0.004$, whereas it is $E(B-V)=1.276\pm0.005$ in a wind medium.
    
    \item[\textbf{(iii)}] The spectral index for the optical and NIR emission within the first 14 days is, $\beta=0.447\pm0.001$ (uniform) and $\beta=0.425\pm0.003$ (wind).
    \item[\textbf{(iv)}] The wind medium model is marginally preferred, based on the fit criteria, however, it cannot reproduce the observed temporal evolution.
    \item[\textbf{(v)}] The uniform medium model can reproduce the temporal evolution of NIR/optical and X-ray data when a non-spreading, edge-effect jet break at early times is invoked.
\end{itemize}
Further, these results have shown:
\begin{itemize}
    \item[\textbf{(vi)}] The accelerated electron index preferred by the early spectral data has $p<2$, with: $p=1.894\pm0.002$ (uniform), and $~p=1.850\pm0.006$ (wind).
    \item[\textbf{(vii)}] The reddening from model fits is consistent with the \citet{schlafly2011} value, $E(B-V) = 1.32$.
    \item[\textbf{(viii)}] Using the \citet{schlafly2011} table value for $A_V/E(B-V) = 2.74$, for the Milky-Way only extinction, our reddening values return $A_{V,~{\rm MW}} \simeq 3.58$ (uniform), and $A_{V,~{\rm MW}} \simeq 3.50$ (wind).
    \item[\textbf{(ix)}] The host contribution to extinction is then, $A_{V,~{\rm host}} \simeq 0.82$ (uniform), and $A_{V,~{\rm host}} \simeq 0.83$ (wind).
    \item[\textbf{(x)}] A jet break at NIR/optical and X-ray is predicted at $\sim 0.5 - 1.0$ days, consistent with a break seen in the optical data.
\end{itemize}
The model fits at four epochs; $\sim$0.5, $\sim$4, $\sim$10, and $\sim$13 days, predict the expected X-ray flux density at 0.3 - 10 keV.
These predictions are consistent with the observed XRT data where the photon index is fixed at $\Gamma = 1.73\pm0.17$ (see Figure\,\ref{fig:combo}).

The $p<2$ value found for NIR/optical/X-ray data within the first 14 days invites questions about the acceleration process and/or magnetic field structure in GRB afterglow shocks.
The required early jet break implies a very narrow jet, while the lack of spreading (edge-effect only) suggests the narrow jet component is surrounded by an energetic sheath, or second component.
In the scenario required to explain the NIR/optical/X-ray temporal and spectral evolution, the radio afterglow will be dominated by emission from the second component -- requiring different microphysical parameters, especially the electron index, which is expected to be $p>2$ (light curve analysis of this model will be shown in Grabham et al. in prep).

Very bright GRBs like the BOAT are extremely valuable, but we still need more of such events (a ``bigger BOAT'') to fully understand GRB afterglows.
Future observations, with spectral and IR coverage are important, as demonstrated here, where the `too blue' afterglow was only identified via the JWST spectral observations \citep{levan2023}.
How this fits in with the population of GRBs is an open question that requires dedicated, high signal-to-noise, NIR/optical/UV spectral coverage of GRB afterglows within the first 10-20 days.
Additionally, theory studies of the magnetic field structure, particle acceleration and simulations of GRB populations aimed at reproducing, not only the BOAT afterglow, but the observed diversity in all (collapsar) GRBs.
Deviations from the classic theory predictions for GRB afterglows, such as the $p<2$ and apparent jet structure requirements for the BOAT afterglow, hold the key to unravelling the physics at work within these high energy phenomena.

\section*{Acknowledgements}

We sincerely thank the anonymous referee for their insightful and helpful
comments that have greatly improved this work. GPL thanks engaging and insightful discussions with Jonathan Granot, Anatoly Spitkovsky, Yuri Sato, Masaomi Tanaka, Kazumi Kashiyama, Nick Ekanger, Riki Matsui, Seiji Toshikage, Riku Kuze, and Koya Chiba.
NMK kindly thanks Masaomi Tanaka and Hamid Hamidani 
for hosting a visit to Tohoku University.
NMK, GPL, and CMBO thank the Yukawa Institute for Theoretical Physics at Kyoto University. 
Discussions during the YITP long-term workshop YITP-T-26-02 on "Multi-Messenger Astrophysics in the Dynamic Universe" were useful to complete this work. NMK acknowledges support from LIV.INNO (STFC Grant ST/W006766/1). GPL and CMBO acknowledge support from the Royal Society (Grant Nos. DHF-R1-221175 and DHF-ERE-221005).
HH acknowledges support from the JSPS Grant-in-Aid for Scientific Research (Grant Nos. 26H00817 and 26K07147).

\section*{Data Availability}

This work is based on observations made with the NASA/ESA/CSA James Webb Space Telescope.
The data were obtained from the Mikulski Archive for Space Telescopes at the Space Telescope Science Institute, which is operated by the Association of Universities for Research in Astronomy, Inc., under NASA contract NAS 5-03127 for JWST. 
These observations are associated with program No. 2782.
This paper is partly based on observations collected at the European Southern Observatory under ESO programme 110.24CF (PI Tanvir).
Data used in this work is available via reasonable request to the authors.



\bibliographystyle{mnras}
\bibliography{ms} 

@ARTICLE{abe2025,
       author = {{Abe}, K. and {Abe}, S. and {Abhishek}, A. and {Acero}, F. and {Aguasca-Cabot}, A. and {Agudo}, I. and {Alispach}, C. and {Ambrosino}, D. and {Ambrosino}, F. and {Antonelli}, L.~A. and {Aramo}, C. and {Arbet-Engels}, A. and {Arcaro}, C. and {Arnesen}, T.~T.~H. and {Asano}, K. and {Aubert}, P. and {Baktash}, A. and {Balbo}, M. and {Bamba}, A. and {Baquero Larriva}, A. and {Barres de Almeida}, U. and {Barrio}, J.~A. and {Barrios Jim{\'e}nez}, L. and {Batkovic}, I. and {Baxter}, J. and {Becerra Gonz{\'a}lez}, J. and {Bernardini}, E. and {Bernete}, J. and {Berti}, A. and {Bezshyiko}, I. and {Bigongiari}, C. and {Bissaldi}, E. and {Blanch}, O. and {Bonnoli}, G. and {Bordas}, P. and {Borkowski}, G. and {Brunelli}, G. and {Bulgarelli}, A. and {Bunse}, M. and {Burelli}, I. and {Burmistrov}, L. and {Cardillo}, M. and {Caroff}, S. and {Carosi}, A. and {Carraro}, R. and {Carrasco}, M.~S. and {Cassol}, F. and {Cerasole}, D. and {Ceribella}, G. and {Cervi{\~n}o Cort{\'\i}nez}, A. and {Chai}, Y. and {Cheng}, K. and {Chiavassa}, A. and {Chikawa}, M. and {Chon}, G. and {Chytka}, L. and {Cicciari}, G.~M. and {Cifuentes}, A. and {Contreras}, J.~L. and {Cortina}, J. and {Costantini}, H. and {Dalchenko}, M. and {Da Vela}, P. and {Dazzi}, F. and {De Angelis}, A. and {de Bony de Lavergne}, M. and {Del Burgo}, R. and {Delgado}, C. and {Delgado Mengual}, J. and {Dellaiera}, M. and {della Volpe}, D. and {De Lotto}, B. and {Del Peral}, L. and {de Menezes}, R. and {De Palma}, G. and {D{\'\i}az}, C. and {Di Piano}, A. and {Di Pierro}, F. and {Di Tria}, R. and {Di Venere}, L. and {Dominik}, R.~M. and {Dominis Prester}, D. and {Donini}, A. and {Dorner}, D. and {Doro}, M. and {Eisenberger}, L. and {Els{\"a}sser}, D. and {Emery}, G. and {Escudero}, J. and {Fallah Ramazani}, V. and {Ferrarotto}, F. and {Fiasson}, A. and {Foffano}, L. and {Fr{\'\i}as Garc{\'\i}a-Lago}, F. and {Fr{\"o}se}, S. and {Fukazawa}, Y. and {Gallozzi}, S. and {Garcia L{\'o}pez}, R. and {Garcia Soto}, S. and {Gasbarra}, C. and {Gasparrini}, D. and {Geyer}, D. and {Giesbrecht Paiva}, J. and {Giglietto}, N. and {Giordano}, F. and {Godinovic}, N. and {Gradetzke}, T. and {Grau}, R. and {Green}, D. and {Green}, J. and {Gunji}, S. and {G{\"u}nther}, P. and {Hackfeld}, J. and {Hadasch}, D. and {Hahn}, A. and {Hashizume}, M. and {Hassan}, T. and {Hayashi}, K. and {Heckmann}, L. and {Heller}, M. and {Herrera Llorente}, J. and {Hirotani}, K. and {Hoffmann}, D. and {Horns}, D. and {Houles}, J. and {Hrabovsky}, M. and {Hrupec}, D. and {Hui}, D. and {Iarlori}, M. and {Imazawa}, R. and {Inada}, T. and {Inome}, Y. and {Inoue}, S. and {Ioka}, K. and {Iori}, M. and {Itokawa}, T. and {Iuliano}, A. and {Jahanvi}, J. and {Jimenez Martinez}, I. and {Jimenez Quiles}, J. and {Jorge Rodrigo}, I. and {Jurysek}, J. and {Kagaya}, M. and {Kalashev}, O. and {Karas}, V. and {Katagiri}, H. and {Kerszberg}, D. and {Kiyomot}, T. and {Kobayashi}, Y. and {Kohri}, K. and {Kong}, A. and {Kornecki}, P. and {Kubo}, H. and {Kushida}, J. and {Lacave}, B. and {Lainez}, M. and {Lamanna}, G. and {Lamastra}, A. and {Lemoigne}, L. and {Linhoff}, M. and {Lombardi}, S. and {Longo}, F. and {L{\'o}pez-Coto}, R. and {L{\'o}pez-Moya}, M. and {L{\'o}pez-Oramas}, A. and {Loporchio}, S. and {Lorini}, A. and {Lozano Bahilo}, J. and {Lucarelli}, F. and {Luciani}, H. and {Luque-Escamilla}, P.~L. and {Majumdar}, P. and {Makariev}, M. and {Mallamaci}, M. and {Mandat}, D. and {Manganaro}, M. and {Maniadakis}, D.~K. and {Manic{\`o}}, G. and {Mannheim}, K. and {Marchesi}, S. and {Marini}, F. and {Mariotti}, M. and {Marquez}, P. and {Marsella}, G. and {Mart{\'\i}}, J. and {Martinez}, O. and {Mart{\'\i}nez}, G. and {Mart{\'\i}nez}, M. and {Mas-Aguilar}, A. and {Massa}, M. and {Maurin}, G. and {Mazin}, D. and {M{\'e}ndez-Gallego}, J. and {Menon}, S. and {Mestre Guillen}, E. and {Miceli}, D. and {Miener}, T. and {Miranda}, J.~M. and {Mirzoyan}, R. and {Mizote}, M.},
        title = "{GRB 221009A: Observations with LST-1 of CTAO and Implications for Structured Jets in Long Gamma-Ray Bursts}",
      journal = {\apjl},
         year = 2025,
        month = jul,
       volume = {988},
       number = {2},
          eid = {L42},
        pages = {L42},
          doi = {10.3847/2041-8213/ade4cf},
archivePrefix = {arXiv},
       eprint = {2507.03077},
 primaryClass = {astro-ph.HE},
       adsurl = {https://ui.adsabs.harvard.edu/abs/2025ApJ...988L..42A}
}

@ARTICLE{achterberg2001,
       author = {{Achterberg}, Abraham and {Gallant}, Yves A. and {Kirk}, John G. and {Guthmann}, Axel W.},
        title = "{Particle acceleration by ultrarelativistic shocks: theory and simulations}",
      journal = {\mnras},
         year = 2001,
        month = dec,
       volume = {328},
       number = {2},
        pages = {393-408},
          doi = {10.1046/j.1365-8711.2001.04851.x},
archivePrefix = {arXiv},
       eprint = {astro-ph/0107530},
 primaryClass = {astro-ph},
       adsurl = {https://ui.adsabs.harvard.edu/abs/2001MNRAS.328..393A}
}

@article{bilby_paper,
    author = "Ashton, Gregory and others",
    title = "{BILBY: A user-friendly Bayesian inference library for gravitational-wave astronomy}",
    eprint = "1811.02042",
    archivePrefix = "arXiv",
    primaryClass = "astro-ph.IM",
    doi = "10.3847/1538-4365/ab06fc",
    journal = "Astrophys. J. Suppl.",
    volume = "241",
    number = "2",
    pages = "27",
    year = "2019"
}

@ARTICLE{blanchard2024,
       author = {{Blanchard}, Peter K. and {Villar}, V. Ashley and {Chornock}, Ryan and {Laskar}, Tanmoy and {Li}, Yijia and {Leja}, Joel and {Pierel}, Justin and {Berger}, Edo and {Margutti}, Raffaella and {Alexander}, Kate D. and {Barnes}, Jennifer and {Cendes}, Yvette and {Eftekhari}, Tarraneh and {Kasen}, Daniel and {LeBaron}, Natalie and {Metzger}, Brian D. and {Muzerolle Page}, James and {Rest}, Armin and {Sears}, Huei and {Siegel}, Daniel M. and {Yadavalli}, S. Karthik},
        title = "{JWST detection of a supernova associated with GRB 221009A without an r-process signature}",
      journal = {Nature Astronomy},
         year = 2024,
        month = jun,
       volume = {8},
        pages = {774-785},
          doi = {10.1038/s41550-024-02237-4},
archivePrefix = {arXiv},
       eprint = {2308.14197},
 primaryClass = {astro-ph.HE},
       adsurl = {https://ui.adsabs.harvard.edu/abs/2024NatAs...8..774B}
}

@ARTICLE{chevalier1999,
       author = {{Chevalier}, Roger A. and {Li}, Zhi-Yun},
        title = "{Gamma-Ray Burst Environments and Progenitors}",
      journal = {\apjl},
         year = 1999,
        month = jul,
       volume = {520},
       number = {1},
        pages = {L29-L32},
          doi = {10.1086/312147},
archivePrefix = {arXiv},
       eprint = {astro-ph/9904417},
 primaryClass = {astro-ph},
       adsurl = {https://ui.adsabs.harvard.edu/abs/1999ApJ...520L..29C}
}

@ARTICLE{fitzpatrick1999,
       author = {{Fitzpatrick}, Edward L.},
        title = "{Correcting for the Effects of Interstellar Extinction}",
      journal = {\pasp},
         year = 1999,
        month = jan,
       volume = {111},
       number = {755},
        pages = {63-75},
          doi = {10.1086/316293},
archivePrefix = {arXiv},
       eprint = {astro-ph/9809387},
 primaryClass = {astro-ph},
       adsurl = {https://ui.adsabs.harvard.edu/abs/1999PASP..111...63F}
}

@ARTICLE{foffano2024,
       author = {{Foffano}, Luca and {Tavani}, Marco and {Piano}, Giovanni},
        title = "{Theoretical Modeling of the Exceptional GRB 221009A Afterglow}",
      journal = {\apjl},
         year = 2024,
        month = oct,
       volume = {973},
       number = {2},
          eid = {L44},
        pages = {L44},
          doi = {10.3847/2041-8213/ad76a3},
archivePrefix = {arXiv},
       eprint = {2409.02859},
 primaryClass = {astro-ph.HE},
       adsurl = {https://ui.adsabs.harvard.edu/abs/2024ApJ...973L..44F}
}

@ARTICLE{fulton2023,
       author = {{Fulton}, M.~D. and {Smartt}, S.~J. and {Rhodes}, L. and {Huber}, M.~E. and {Villar}, V.~A. and {Moore}, T. and {Srivastav}, S. and {Schultz}, A.~S.~B. and {Chambers}, K.~C. and {Izzo}, L. and {Hjorth}, J. and {Chen}, T. -W. and {Nicholl}, M. and {Foley}, R.~J. and {Rest}, A. and {Smith}, K.~W. and {Young}, D.~R. and {Sim}, S.~A. and {Bright}, J. and {Zenati}, Y. and {de Boer}, T. and {Bulger}, J. and {Fairlamb}, J. and {Gao}, H. and {Lin}, C. -C. and {Lowe}, T. and {Magnier}, E.~A. and {Smith}, I.~A. and {Wainscoat}, R. and {Coulter}, D.~A. and {Jones}, D.~O. and {Kilpatrick}, C.~D. and {McGill}, P. and {Ramirez-Ruiz}, E. and {Lee}, K. -S. and {Narayan}, G. and {Ramakrishnan}, V. and {Ridden-Harper}, R. and {Singh}, A. and {Wang}, Q. and {Kong}, A.~K.~H. and {Ngeow}, C. -C. and {Pan}, Y. -C. and {Yang}, S. and {Davis}, K.~W. and {Piro}, A.~L. and {Rojas-Bravo}, C. and {Sommer}, J. and {Yadavalli}, S.~K.},
        title = "{The Optical Light Curve of GRB 221009A: The Afterglow and the Emerging Supernova}",
      journal = {\apjl},
         year = 2023,
        month = mar,
       volume = {946},
       number = {1},
          eid = {L22},
        pages = {L22},
          doi = {10.3847/2041-8213/acc101},
archivePrefix = {arXiv},
       eprint = {2301.11170},
 primaryClass = {astro-ph.HE},
       adsurl = {https://ui.adsabs.harvard.edu/abs/2023ApJ...946L..22F}
}

@ARTICLE{gao2013,
       author = {{Gao}, He and {Lei}, Wei-Hua and {Zou}, Yuan-Chuan and {Wu}, Xue-Feng and {Zhang}, Bing},
        title = "{A complete reference of the analytical synchrotron external shock models of gamma-ray bursts}",
      journal = {\nar},
         year = 2013,
        month = dec,
       volume = {57},
       number = {6},
        pages = {141-190},
          doi = {10.1016/j.newar.2013.10.001},
archivePrefix = {arXiv},
       eprint = {1310.2181},
 primaryClass = {astro-ph.HE},
       adsurl = {https://ui.adsabs.harvard.edu/abs/2013NewAR..57..141G}
}

@ARTICLE{granot2002,
       author = {{Granot}, Jonathan and {Sari}, Re'em},
        title = "{The Shape of Spectral Breaks in Gamma-Ray Burst Afterglows}",
      journal = {\apj},
         year = 2002,
        month = apr,
       volume = {568},
       number = {2},
        pages = {820-829},
          doi = {10.1086/338966},
archivePrefix = {arXiv},
       eprint = {astro-ph/0108027},
 primaryClass = {astro-ph},
       adsurl = {https://ui.adsabs.harvard.edu/abs/2002ApJ...568..820G}
}

@ARTICLE{kann2010,
       author = {{Kann}, D.~A. and {Klose}, S. and {Zhang}, B. and {Malesani}, D. and {Nakar}, E. and {Pozanenko}, A. and {Wilson}, A.~C. and {Butler}, N.~R. and {Jakobsson}, P. and {Schulze}, S. and {Andreev}, M. and {Antonelli}, L.~A. and {Bikmaev}, I.~F. and {Biryukov}, V. and {B{\"o}ttcher}, M. and {Burenin}, R.~A. and {Castro Cer{\'o}n}, J.~M. and {Castro-Tirado}, A.~J. and {Chincarini}, G. and {Cobb}, B.~E. and {Covino}, S. and {D'Avanzo}, P. and {D'Elia}, V. and {Della Valle}, M. and {de Ugarte Postigo}, A. and {Efimov}, Yu. and {Ferrero}, P. and {Fugazza}, D. and {Fynbo}, J.~P.~U. and {G{\r{a}}lfalk}, M. and {Grundahl}, F. and {Gorosabel}, J. and {Gupta}, S. and {Guziy}, S. and {Hafizov}, B. and {Hjorth}, J. and {Holhjem}, K. and {Ibrahimov}, M. and {Im}, M. and {Israel}, G.~L. and {Je{\'l}inek}, M. and {Jensen}, B.~L. and {Karimov}, R. and {Khamitov}, I.~M. and {Kizilo{\v{g}}lu}, {\"U}. and {Klunko}, E. and {Kub{\'a}nek}, P. and {Kutyrev}, A.~S. and {Laursen}, P. and {Levan}, A.~J. and {Mannucci}, F. and {Martin}, C.~M. and {Mescheryakov}, A. and {Mirabal}, N. and {Norris}, J.~P. and {Ovaldsen}, J. -E. and {Paraficz}, D. and {Pavlenko}, E. and {Piranomonte}, S. and {Rossi}, A. and {Rumyantsev}, V. and {Salinas}, R. and {Sergeev}, A. and {Sharapov}, D. and {Sollerman}, J. and {Stecklum}, B. and {Stella}, L. and {Tagliaferri}, G. and {Tanvir}, N.~R. and {Telting}, J. and {Testa}, V. and {Updike}, A.~C. and {Volnova}, A. and {Watson}, D. and {Wiersema}, K. and {Xu}, D.},
        title = "{The Afterglows of Swift-era Gamma-ray Bursts. I. Comparing pre-Swift and Swift-era Long/Soft (Type II) GRB Optical Afterglows}",
      journal = {\apj},
         year = 2010,
        month = sep,
       volume = {720},
       number = {2},
        pages = {1513-1558},
          doi = {10.1088/0004-637X/720/2/1513},
archivePrefix = {arXiv},
       eprint = {0712.2186},
 primaryClass = {astro-ph},
       adsurl = {https://ui.adsabs.harvard.edu/abs/2010ApJ...720.1513K}
}

@ARTICLE{kann2023,
       author = {{Kann}, D.~A. and {Agayeva}, S. and {Aivazyan}, V. and {Alishov}, S. and {Andrade}, C.~M. and {Antier}, S. and {Baransky}, A. and {Bendjoya}, P. and {Benkhaldoun}, Z. and {Beradze}, S. and {Berezin}, D. and {Bo{\"e}r}, M. and {Broens}, E. and {Brunier}, S. and {Bulla}, M. and {Burkhonov}, O. and {Burns}, E. and {Chen}, Y. and {Chen}, Y.~P. and {Conti}, M. and {Coughlin}, M.~W. and {Cui}, W.~W. and {Daigne}, F. and {Delaveau}, B. and {Devillepoix}, H.~A.~R. and {Dietrich}, T. and {Dornic}, D. and {Dubois}, F. and {Ducoin}, J. -G. and {Durand}, E. and {Duverne}, P. -A. and {Eggenstein}, H. -B. and {Ehgamberdiev}, S. and {Fouad}, A. and {Freeberg}, M. and {Froebrich}, D. and {Ge}, M.~Y. and {Gervasoni}, S. and {Godunova}, V. and {Gokuldass}, P. and {Gurbanov}, E. and {Han}, D.~W. and {Hasanov}, E. and {Hello}, P. and {Hussenot-Desenonges}, T. and {Inasaridze}, R. and {Iskandar}, A. and {Ismailov}, N. and {Janati}, A. and {du Laz}, T. Jegou and {Jia}, S.~M. and {Karpov}, S. and {Kaeouach}, A. and {Kiendrebeogo}, R.~W. and {Klotz}, A. and {Kneip}, R. and {Kochiashvili}, N. and {Kunert}, N. and {Lekic}, A. and {Leonini}, S. and {Li}, C.~K. and {Li}, W. and {Li}, X.~B. and {Liao}, J.~Y. and {Logie}, L. and {Lu}, F.~J. and {Mao}, J. and {Marchais}, D. and {M{\'e}nard}, R. and {Morris}, D. and {Natsvlishvili}, R. and {Nedora}, V. and {Noonan}, K. and {Noysena}, K. and {Orange}, N.~B. and {Pang}, P.~T.~H. and {Peng}, H.~W. and {Pellouin}, C. and {Peloton}, J. and {Pradier}, T. and {Pyshna}, O. and {Rajabov}, Y. and {Rau}, S. and {Rinner}, C. and {Rivet}, J. -P. and {Romanov}, F.~D. and {Rosi}, P. and {Rupchandani}, V.~A. and {Serrau}, M. and {Shokry}, A. and {Simon}, A. and {Smith}, K. and {Sokoliuk}, O. and {Soliman}, M. and {Song}, L.~M. and {Takey}, A. and {Tillayev}, Y. and {Ramirez}, L.~M. Tinjaca and {e Melo}, I. Tosta and {Turpin}, D. and {de Ugarte Postigo}, A. and {Vanaverbeke}, S. and {Vasylenko}, V. and {Vernet}, D. and {Vidadi}, Z. and {Wang}, C. and {Wang}, J. and {Wang}, L.~T. and {Wang}, X.~F. and {Xiong}, S.~L. and {Xu}, Y.~P. and {Xue}, W.~C. and {Zeng}, X. and {Zhang}, S.~N. and {Zhao}, H.~S. and {Zhao}, X.~F.},
        title = "{GRANDMA and HXMT Observations of GRB 221009A: The Standard Luminosity Afterglow of a Hyperluminous Gamma-Ray Burst-In Gedenken an David Alexander Kann}",
      journal = {\apjl},
         year = 2023,
        month = may,
       volume = {948},
       number = {2},
          eid = {L12},
        pages = {L12},
          doi = {10.3847/2041-8213/acc8d0},
archivePrefix = {arXiv},
       eprint = {2302.06225},
 primaryClass = {astro-ph.HE},
       adsurl = {https://ui.adsabs.harvard.edu/abs/2023ApJ...948L..12K}
}

@ARTICLE{keshet2005,
       author = {{Keshet}, Uri and {Waxman}, Eli},
        title = "{Energy Spectrum of Particles Accelerated in Relativistic Collisionless Shocks}",
      journal = {\prl},
         year = 2005,
        month = mar,
       volume = {94},
       number = {11},
          eid = {111102},
        pages = {111102},
          doi = {10.1103/PhysRevLett.94.111102},
archivePrefix = {arXiv},
       eprint = {astro-ph/0408489},
 primaryClass = {astro-ph},
       adsurl = {https://ui.adsabs.harvard.edu/abs/2005PhRvL..94k1102K}
}

@ARTICLE{klinger2024,
       author = {{Klinger}, Marc and {Taylor}, Andrew M. and {Parsotan}, Tyler and {Beardmore}, Andrew and {Heinz}, Sebastian and {Zhu}, Sylvia J.},
        title = "{The multiwavelength picture of GRB 221009A's afterglow}",
      journal = {\mnras},
         year = 2024,
        month = mar,
       volume = {529},
       number = {1},
        pages = {L47-L53},
          doi = {10.1093/mnrasl/slad185},
archivePrefix = {arXiv},
       eprint = {2308.13854},
 primaryClass = {astro-ph.HE},
       adsurl = {https://ui.adsabs.harvard.edu/abs/2024MNRAS.529L..47K}
}

@ARTICLE{kong2024,
       author = {{Kong}, De-Feng and {Wang}, Xiang-Gao and {Zheng}, WeiKang and {L{\"u}}, Hou-Jun and {Xin}, L.~P. and {Lin}, Da-Bin and {Cao}, Jia-Xin and {Lu}, Ming-Xuan and {Ren}, B. and {Vidal}, Edgar P. and {Wei}, J.~Y. and {Liang}, En-Wei and {Filippenko}, Alexei V.},
        title = "{GRB 221009A/SN 2022xiw: A Supernova Obscured by a Gamma-Ray Burst Afterglow?}",
      journal = {\apj},
         year = 2024,
        month = aug,
       volume = {971},
       number = {1},
          eid = {56},
        pages = {56},
          doi = {10.3847/1538-4357/ad5ce1},
archivePrefix = {arXiv},
       eprint = {2407.00639},
 primaryClass = {astro-ph.HE},
       adsurl = {https://ui.adsabs.harvard.edu/abs/2024ApJ...971...56K}
}

@ARTICLE{lamb2021,
       author = {{Lamb}, Gavin P. and {Kann}, D. Alexander and {Fern{\'a}ndez}, Joseph John and {Mandel}, Ilya and {Levan}, Andrew J. and {Tanvir}, Nial R.},
        title = "{GRB jet structure and the jet break}",
      journal = {\mnras},
         year = 2021,
        month = sep,
       volume = {506},
       number = {3},
        pages = {4163-4174},
          doi = {10.1093/mnras/stab2071},
archivePrefix = {arXiv},
       eprint = {2104.11099},
 primaryClass = {astro-ph.HE},
       adsurl = {https://ui.adsabs.harvard.edu/abs/2021MNRAS.506.4163L}
}

@ARTICLE{laskar2023,
       author = {{Laskar}, Tanmoy and {Alexander}, Kate D. and {Margutti}, Raffaella and {Eftekhari}, Tarraneh and {Chornock}, Ryan and {Berger}, Edo and {Cendes}, Yvette and {Duerr}, Anne and {Perley}, Daniel A. and {Ravasio}, Maria Edvige and {Yamazaki}, Ryo and {Ayache}, Eliot H. and {Barclay}, Thomas and {Duran}, Rodolfo Barniol and {Bhandari}, Shivani and {Brethauer}, Daniel and {Christy}, Collin T. and {Coppejans}, Deanne L. and {Duffell}, Paul and {Fong}, Wen-fai and {Gomboc}, Andreja and {Guidorzi}, Cristiano and {Kennea}, Jamie A. and {Kobayashi}, Shiho and {Levan}, Andrew and {Lobanov}, Andrei P. and {Metzger}, Brian D. and {Ros}, Eduardo and {Schroeder}, Genevieve and {Williams}, P.~K.~G.},
        title = "{The Radio to GeV Afterglow of GRB 221009A}",
      journal = {\apjl},
         year = 2023,
        month = mar,
       volume = {946},
       number = {1},
          eid = {L23},
        pages = {L23},
          doi = {10.3847/2041-8213/acbfad},
archivePrefix = {arXiv},
       eprint = {2302.04388},
 primaryClass = {astro-ph.HE},
       adsurl = {https://ui.adsabs.harvard.edu/abs/2023ApJ...946L..23L}
}

@ARTICLE{lesage2023,
       author = {{Lesage}, S. and {Veres}, P. and {Briggs}, M.~S. and {Goldstein}, A. and {Kocevski}, D. and {Burns}, E. and {Wilson-Hodge}, C.~A. and {Bhat}, P.~N. and {Huppenkothen}, D. and {Fryer}, C.~L. and {Hamburg}, R. and {Racusin}, J. and {Bissaldi}, E. and {Cleveland}, W.~H. and {Dalessi}, S. and {Fletcher}, C. and {Giles}, M.~M. and {Hristov}, B.~A. and {Hui}, C.~M. and {Mailyan}, B. and {Malacaria}, C. and {Poolakkil}, S. and {Roberts}, O.~J. and {von Kienlin}, A. and {Wood}, J. and {Ajello}, M. and {Arimoto}, M. and {Baldini}, L. and {Ballet}, J. and {Baring}, M.~G. and {Bastieri}, D. and {Gonzalez}, J. Becerra and {Bellazzini}, R. and {Bissaldi}, E. and {Blandford}, R.~D. and {Bonino}, R. and {Bruel}, P. and {Buson}, S. and {Cameron}, R.~A. and {Caputo}, R. and {Caraveo}, P.~A. and {Cavazzuti}, E. and {Chiaro}, G. and {Cibrario}, N. and {Ciprini}, S. and {Orestano}, P. Cristarella and {Crnogorcevic}, M. and {Cuoco}, A. and {Cutini}, S. and {D'Ammando}, F. and {De Gaetano}, S. and {Di Lalla}, N. and {Di Venere}, L. and {Dom{\'\i}nguez}, A. and {Fegan}, S.~J. and {Ferrara}, E.~C. and {Fleischhack}, H. and {Fukazawa}, Y. and {Funk}, S. and {Fusco}, P. and {Galanti}, G. and {Gammaldi}, V. and {Gargano}, F. and {Gasbarra}, C. and {Gasparrini}, D. and {Germani}, S. and {Giacchino}, F. and {Giglietto}, N. and {Gill}, R. and {Giroletti}, M. and {Granot}, J. and {Green}, D. and {Grenier}, I.~A. and {Guiriec}, S. and {Gustafsson}, M. and {Hays}, E. and {Hewitt}, J.~W. and {Horan}, D. and {Hou}, X. and {Kuss}, M. and {Latronico}, L. and {Laviron}, A. and {Lemoine-Goumard}, M. and {Li}, J. and {Liodakis}, I. and {Longo}, F. and {Loparco}, F. and {Lorusso}, L. and {Lovellette}, M.~N. and {Lubrano}, P. and {Maldera}, S. and {Manfreda}, A. and {Mart{\'\i}-Devesa}, G. and {Mazziotta}, M.~N. and {McEnery}, J.~E. and {Mereu}, I. and {Meyer}, M. and {Michelson}, P.~F. and {Mizuno}, T. and {Monzani}, M.~E. and {Morselli}, A. and {Moskalenko}, I.~V. and {Negro}, M. and {Nuss}, E. and {Omodei}, N. and {Orlando}, E. and {Ormes}, J.~F. and {Paneque}, D. and {Panzarini}, G. and {Persic}, M. and {Pesce-Rollins}, M. and {Pillera}, R. and {Piron}, F. and {Poon}, H. and {Porter}, T.~A. and {Principe}, G. and {Rain{\`o}}, S. and {Rando}, R. and {Rani}, B. and {Razzano}, M. and {Razzaque}, S. and {Reimer}, A. and {Reimer}, O. and {Ryde}, F. and {S{\'a}nchez-Conde}, M. and {Parkinson}, P.~M. Saz and {Scotton}, L. and {Serini}, D. and {Sgr{\`o}}, C. and {Sharma}, V. and {Siskind}, E.~J. and {Spandre}, G. and {Spinelli}, P. and {Tajima}, H. and {Torres}, D.~F. and {Valverde}, J. and {Venters}, T. and {Wadiasingh}, Z. and {Wood}, K. and {Zaharijas}, G.},
        title = "{Fermi-GBM Discovery of GRB 221009A: An Extraordinarily Bright GRB from Onset to Afterglow}",
      journal = {\apjl},
         year = 2023,
        month = aug,
       volume = {952},
       number = {2},
          eid = {L42},
        pages = {L42},
          doi = {10.3847/2041-8213/ace5b4},
archivePrefix = {arXiv},
       eprint = {2303.14172},
 primaryClass = {astro-ph.HE},
       adsurl = {https://ui.adsabs.harvard.edu/abs/2023ApJ...952L..42L}
}

@ARTICLE{levan2023,
       author = {{Levan}, A.~J. and {Lamb}, G.~P. and {Schneider}, B. and {Hjorth}, J. and {Zafar}, T. and {de Ugarte Postigo}, A. and {Sargent}, B. and {Mullally}, S.~E. and {Izzo}, L. and {D'Avanzo}, P. and {Burns}, E. and {Ag{\"u}{\'\i} Fern{\'a}ndez}, J.~F. and {Barclay}, T. and {Bernardini}, M.~G. and {Bhirombhakdi}, K. and {Bremer}, M. and {Brivio}, R. and {Campana}, S. and {Chrimes}, A.~A. and {D'Elia}, V. and {Della Valle}, M. and {De Pasquale}, M. and {Ferro}, M. and {Fong}, W. and {Fruchter}, A.~S. and {Fynbo}, J.~P.~U. and {Gaspari}, N. and {Gompertz}, B.~P. and {Hartmann}, D.~H. and {Hedges}, C.~L. and {Heintz}, K.~E. and {Hotokezaka}, K. and {Jakobsson}, P. and {Kann}, D.~A. and {Kennea}, J.~A. and {Laskar}, T. and {Le Floc'h}, E. and {Malesani}, D.~B. and {Melandri}, A. and {Metzger}, B.~D. and {Oates}, S.~R. and {Pian}, E. and {Piranomonte}, S. and {Pugliese}, G. and {Racusin}, J.~L. and {Rastinejad}, J.~C. and {Ravasio}, M.~E. and {Rossi}, A. and {Saccardi}, A. and {Salvaterra}, R. and {Sbarufatti}, B. and {Starling}, R.~L.~C. and {Tanvir}, N.~R. and {Th{\"o}ne}, C.~C. and {van der Horst}, A.~J. and {Vergani}, S.~D. and {Watson}, D. and {Wiersema}, K. and {Wijers}, R.~A.~M.~J. and {Xu}, Dong},
        title = "{The First JWST Spectrum of a GRB Afterglow: No Bright Supernova in Observations of the Brightest GRB of all Time, GRB 221009A}",
      journal = {\apjl},
         year = 2023,
        month = mar,
       volume = {946},
       number = {1},
          eid = {L28},
        pages = {L28},
          doi = {10.3847/2041-8213/acc2c1},
archivePrefix = {arXiv},
       eprint = {2302.07761},
 primaryClass = {astro-ph.HE},
       adsurl = {https://ui.adsabs.harvard.edu/abs/2023ApJ...946L..28L}
}

@ARTICLE{LHAASO2023,
       author = {{LHAASO Collaboration} and {Cao}, Z. and {Aharonian}, F. and {An}, Q. and {Axikegu}, A. and {Bai}, L.~X. and {Bai}, Y.~X. and {Bao}, Y.~W. and {Bastieri}, D. and {Bi}, X.~J. and {Bi}, Y.~J. and {Cai}, J.~T. and {Cao}, Q. and {Cao}, W.~Y. and {Cao}, Z. and {Chang}, J. and {Chang}, J.~F. and {Chen}, E.~S. and {Chen}, L. and {Chen}, L. and {Chen}, L. and {Chen}, M.~J. and {Chen}, M.~L. and {Chen}, Q.~H. and {Chen}, S.~H. and {Chen}, S.~Z. and {Chen}, T.~L. and {Chen}, Y. and {Cheng}, H.~L. and {Cheng}, N. and {Cheng}, Y.~D. and {Cui}, S.~W. and {Cui}, X.~H. and {Cui}, Y.~D. and {Dai}, B.~Z. and {Dai}, H.~L. and {Danzengluobu}, D. and {Della Volpe}, D. and {Dong}, X.~Q. and {Duan}, K.~K. and {Fan}, J.~H. and {Fan}, Y.~Z. and {Fang}, J. and {Fang}, K. and {Feng}, C.~F. and {Feng}, L. and {Feng}, S.~H. and {Feng}, X.~T. and {Feng}, Y.~L. and {Gao}, B. and {Gao}, C.~D. and {Gao}, L.~Q. and {Gao}, Q. and {Gao}, W. and {Gao}, W.~K. and {Ge}, M.~M. and {Geng}, L.~S. and {Gong}, G.~H. and {Gou}, Q.~B. and {Gu}, M.~H. and {Guo}, F.~L. and {Guo}, X.~L. and {Guo}, Y.~Q. and {Guo}, Y.~Y. and {Han}, Y.~A. and {He}, H.~H. and {He}, H.~N. and {He}, J.~Y. and {He}, X.~B. and {He}, Y. and {Heller}, M. and {Hor}, Y.~K. and {Hou}, B.~W. and {Hou}, C. and {Hou}, X. and {Hu}, H.~B. and {Hu}, Q. and {Hu}, S.~C. and {Huang}, D.~H. and {Huang}, T.~Q. and {Huang}, W.~J. and {Huang}, X.~T. and {Huang}, Z.~C. and {Ji}, X.~L. and {Jia}, H.~Y. and {Jia}, K. and {Jiang}, K. and {Jiang}, X.~W. and {Jiang}, Z.~J. and {Jin}, M. and {Kang}, M.~M. and {Ke}, T. and {Kuleshov}, D. and {Kurinov}, K. and {Li}, B.~B. and {Li}, C. and {Li}, C. and {Li}, D. and {Li}, F. and {Li}, H.~B. and {Li}, H.~C. and {Li}, H.~Y. and {Li}, J. and {Li}, J. and {Li}, J. and {Li}, K. and {Li}, W.~L. and {Li}, W.~L. and {Li}, X.~R. and {Li}, X. and {Li}, Y.~Z. and {Li}, Z. and {Li}, Z. and {Liang}, E.~W. and {Liang}, Y.~F. and {Lin}, S.~J. and {Liu}, B. and {Liu}, C. and {Liu}, D. and {Liu}, H. and {Liu}, H.~D. and {Liu}, J. and {Liu}, J.~L. and {Liu}, J.~L. and {Liu}, J.~S. and {Liu}, J.~Y. and {Liu}, M.~Y. and {Liu}, R.~Y. and {Liu}, S.~M. and {Liu}, W. and {Liu}, Y. and {Liu}, Y.~N. and {Long}, W.~J. and {Lu}, R. and {Luo}, Q. and {Lv}, H.~K. and {Ma}, B.~Q. and {Ma}, L.~L. and {Ma}, X.~H. and {Mao}, J.~R. and {Min}, Z. and {Mitthumsiri}, W. and {Nan}, Y.~C. and {Ou}, Z.~W. and {Pang}, B.~Y. and {Pattarakijwanich}, P. and {Pei}, Z.~Y. and {Qi}, M.~Y. and {Qi}, Y.~Q. and {Qiao}, B.~Q. and {Qin}, J.~J. and {Ruffolo}, D. and {Saiz}, A. and {Shao}, C.~Y. and {Shao}, L. and {Shchegolev}, O. and {Sheng}, X.~D. and {Song}, H.~C. and {Stenkin}, Y.~V. and {Stepanov}, V. and {Su}, Y. and {Sun}, Q.~N. and {Sun}, X.~N. and {Sun}, Z.~B. and {Tam}, P.~H.~T. and {Tang}, Z.~B. and {Tian}, W.~W. and {Wang}, C. and {Wang}, C.~B. and {Wang}, G.~W. and {Wang}, H.~G. and {Wang}, H.~H. and {Wang}, J.~C. and {Wang}, J.~S. and {Wang}, K. and {Wang}, L.~P. and {Wang}, L.~Y. and {Wang}, P.~H. and {Wang}, R. and {Wang}, W. and {Wang}, X.~G. and {Wang}, Y.~D. and {Wang}, Y.~J. and {Wang}, Z.~H. and {Wang}, Z.~X. and {Wang}, Z. and {Wei}, D.~M. and {Wei}, J.~J. and {Wei}, Y.~J. and {Wen}, T. and {Wu}, C.~Y. and {Wu}, H.~R. and {Wu}, S. and {Wu}, X.~F. and {Wu}, Y.~S. and {Xi}, S.~Q. and {Xia}, J. and {Xia}, J.~J. and {Xiang}, G.~M. and {Xiao}, D.~X.},
        title = "{A tera-electron volt afterglow from a narrow jet in an extremely bright gamma-ray burst.}",
      journal = {Science},
         year = 2023,
        month = jun,
       volume = {380},
       number = {6652},
        pages = {1390-1396},
          doi = {10.1126/science.adg9328},
archivePrefix = {arXiv},
       eprint = {2306.06372},
 primaryClass = {astro-ph.HE},
       adsurl = {https://ui.adsabs.harvard.edu/abs/2023Sci...380.1390L}
}

@ARTICLE{li2026,
       author = {{Li}, Xiao-Yan and {Liu}, Tong and {Huang}, Bao-Quan and {Deng}, Chen},
        title = "{Statistical analysis of multi-band plateaus in gamma-ray burst afterglows}",
      journal = {arXiv e-prints},
         year = 2026,
        month = jan,
          eid = {arXiv:2601.01586},
        pages = {arXiv:2601.01586},
          doi = {10.48550/arXiv.2601.01586},
archivePrefix = {arXiv},
       eprint = {2601.01586},
 primaryClass = {astro-ph.HE},
       adsurl = {https://ui.adsabs.harvard.edu/abs/2026arXiv260101586L}
}

@ARTICLE{malesani2023,
       author = {{Malesani}, D.~B. and {Levan}, A.~J. and {Izzo}, L. and {de Ugarte Postigo}, A. and {Ghirlanda}, G. and {Heintz}, K.~E. and {Kann}, D.~A. and {Lamb}, G.~P. and {Palmerio}, J. and {Salafia}, O.~S. and {Salvaterra}, R. and {Tanvir}, N.~R. and {Ag{\"u}{\'\i} Fern{\'a}ndez}, J.~F. and {Campana}, S. and {Chrimes}, A.~A. and {D'Avanzo}, P. and {D'Elia}, V. and {Della Valle}, M. and {De Pasquale}, M. and {Fynbo}, J.~P.~U. and {Gaspari}, N. and {Gompertz}, B.~P. and {Hartmann}, D.~H. and {Hjorth}, J. and {Jakobsson}, P. and {Palazzi}, E. and {Pian}, E. and {Pugliese}, G. and {Ravasio}, M.~E. and {Rossi}, A. and {Saccardi}, A. and {Schady}, P. and {Schneider}, B. and {Sollerman}, J. and {Starling}, R.~L.~C. and {Th{\"o}ne}, C.~C. and {van der Horst}, A.~J. and {Vergani}, S.~D. and {Watson}, D. and {Wiersema}, K. and {Xu}, D. and {Zafar}, T. and {Zheng}, S.~Y.},
        title = "{The brightest GRB ever detected: GRB 221009A as a highly luminous event at z = 0.151}",
      journal = {\aap},
         year = 2025,
        month = sep,
       volume = {701},
          eid = {A134},
        pages = {A134},
          doi = {10.1051/0004-6361/202346146},
archivePrefix = {arXiv},
       eprint = {2302.07891},
 primaryClass = {astro-ph.HE},
       adsurl = {https://ui.adsabs.harvard.edu/abs/2025A&A...701A.134M}
}

@ARTICLE{malesani2025,
       author = {{Malesani}, D.~B. and {Levan}, A.~J. and {Izzo}, L. and {de Ugarte Postigo}, A. and {Ghirlanda}, G. and {Heintz}, K.~E. and {Kann}, D.~A. and {Lamb}, G.~P. and {Palmerio}, J. and {Salafia}, O.~S. and {Salvaterra}, R. and {Tanvir}, N.~R. and {Ag{\"u}{\'\i} Fern{\'a}ndez}, J.~F. and {Campana}, S. and {Chrimes}, A.~A. and {D'Avanzo}, P. and {D'Elia}, V. and {Della Valle}, M. and {De Pasquale}, M. and {Fynbo}, J.~P.~U. and {Gaspari}, N. and {Gompertz}, B.~P. and {Hartmann}, D.~H. and {Hjorth}, J. and {Jakobsson}, P. and {Palazzi}, E. and {Pian}, E. and {Pugliese}, G. and {Ravasio}, M.~E. and {Rossi}, A. and {Saccardi}, A. and {Schady}, P. and {Schneider}, B. and {Sollerman}, J. and {Starling}, R.~L.~C. and {Th{\"o}ne}, C.~C. and {van der Horst}, A.~J. and {Vergani}, S.~D. and {Watson}, D. and {Wiersema}, K. and {Xu}, D. and {Zafar}, T. and {Zheng}, S.~Y.},
        title = "{The brightest GRB ever detected: GRB 221009A as a highly luminous event at z = 0.151}",
      journal = {\aap},
         year = 2025,
        month = sep,
       volume = {701},
          eid = {A134},
        pages = {A134},
          doi = {10.1051/0004-6361/202346146},
archivePrefix = {arXiv},
       eprint = {2302.07891},
 primaryClass = {astro-ph.HE},
       adsurl = {https://ui.adsabs.harvard.edu/abs/2025A&A...701A.134M}
}

@ARTICLE{oconnor2023,
       author = {{O'Connor}, Brendan and {Troja}, Eleonora and {Ryan}, Geoffrey and {Beniamini}, Paz and {van Eerten}, Hendrik and {Granot}, Jonathan and {Dichiara}, Simone and {Ricci}, Roberto and {Lipunov}, Vladimir and {Gillanders}, James H. and {Gill}, Ramandeep and {Moss}, Michael and {Anand}, Shreya and {Andreoni}, Igor and {Becerra}, Rosa L. and {Buckley}, David A.~H. and {Butler}, Nathaniel R. and {Cenko}, Stephen B. and {Chasovnikov}, Aristarkh and {Durbak}, Joseph and {Francile}, Carlos and {Hammerstein}, Erica and {van der Horst}, Alexander J. and {Kasliwal}, Mansi M. and {Kouveliotou}, Chryssa and {Kutyrev}, Alexander S. and {Lee}, William H. and {Srinivasaragavan}, Gokul P. and {Topolev}, Vladislav and {Watson}, Alan M. and {Yang}, Yuhan and {Zhirkov}, Kirill},
        title = "{A structured jet explains the extreme GRB 221009A}",
      journal = {Science Advances},
         year = 2023,
        month = jun,
       volume = {9},
       number = {23},
          eid = {eadi1405},
        pages = {eadi1405},
          doi = {10.1126/sciadv.adi1405},
archivePrefix = {arXiv},
       eprint = {2302.07906},
 primaryClass = {astro-ph.HE},
       adsurl = {https://ui.adsabs.harvard.edu/abs/2023SciA....9I1405O}
}

@ARTICLE{rastinejad2024,
       author = {{Rastinejad}, J.~C. and {Fong}, W. and {Levan}, A.~J. and {Tanvir}, N.~R. and {Kilpatrick}, C.~D. and {Fruchter}, A.~S. and {Anand}, S. and {Bhirombhakdi}, K. and {Covino}, S. and {Fynbo}, J.~P.~U. and {Halevi}, G. and {Hartmann}, D.~H. and {Heintz}, K.~E. and {Izzo}, L. and {Jakobsson}, P. and {Kangas}, T. and {Lamb}, G.~P. and {Malesani}, D.~B. and {Melandri}, A. and {Metzger}, B.~D. and {Milvang-Jensen}, B. and {Pian}, E. and {Pugliese}, G. and {Rossi}, A. and {Siegel}, D.~M. and {Singh}, P. and {Stratta}, G.},
        title = "{A Hubble Space Telescope Search for r-Process Nucleosynthesis in Gamma-Ray Burst Supernovae}",
      journal = {\apj},
         year = 2024,
        month = jun,
       volume = {968},
       number = {1},
          eid = {14},
        pages = {14},
          doi = {10.3847/1538-4357/ad409c},
archivePrefix = {arXiv},
       eprint = {2312.04630},
 primaryClass = {astro-ph.HE},
       adsurl = {https://ui.adsabs.harvard.edu/abs/2024ApJ...968...14R}
}

@ARTICLE{ren2024,
       author = {{Ren}, Jia and {Wang}, Yun and {Dai}, Zi-Gao},
        title = "{Jet Structure and Burst Environment of GRB 221009A}",
      journal = {\apj},
         year = 2024,
        month = feb,
       volume = {962},
       number = {2},
          eid = {115},
        pages = {115},
          doi = {10.3847/1538-4357/ad1bcd},
archivePrefix = {arXiv},
       eprint = {2310.15886},
 primaryClass = {astro-ph.HE},
       adsurl = {https://ui.adsabs.harvard.edu/abs/2024ApJ...962..115R}
}

@ARTICLE{rhodes2024,
       author = {{Rhodes}, L. and {van der Horst}, A.~J. and {Bright}, J.~S. and {Leung}, J.~K. and {Anderson}, G.~E. and {Fender}, R. and {Ag{\"u}{\'\i} Fern{\'a}ndez}, J.~F. and {Bremer}, M. and {Chandra}, P. and {Dobie}, D. and {Farah}, W. and {Giarratana}, S. and {Gourdji}, K. and {Green}, D.~A. and {Lenc}, E. and {Micha{\l}owski}, M.~J. and {Murphy}, T. and {Nayana}, A.~J. and {Pollak}, A.~W. and {Rowlinson}, A. and {Schussler}, F. and {Siemion}, A. and {Starling}, R.~L.~C. and {Scott}, P. and {Th{\"o}ne}, C.~C. and {Titterington}, D. and {de Ugarte Postigo}, A.},
        title = "{Rocking the BOAT: the ups and downs of the long-term radio light curve for GRB 221009A}",
      journal = {\mnras},
         year = 2024,
        month = oct,
       volume = {533},
       number = {4},
        pages = {4435-4449},
          doi = {10.1093/mnras/stae2050},
archivePrefix = {arXiv},
       eprint = {2408.16637},
 primaryClass = {astro-ph.HE},
       adsurl = {https://ui.adsabs.harvard.edu/abs/2024MNRAS.533.4435R}
}

@ARTICLE{sanchez-ramirez2024,
       author = {{S{\'a}nchez-Ram{\'\i}rez}, R. and {Lang}, R.~G. and {Pozanenko}, A. and {Mart{\'\i}nez-Huerta}, H. and {Hu}, Y. -D. and {Pandey}, S.~B. and {Gupta}, R. and {Ror}, A.~K. and {Zhang}, B. -B. and {Caballero-Garc{\'\i}a}, M.~D. and {Oates}, S.~R. and {P{\'e}rez-Garc{\'\i}a}, I. and {Guziy}, S. and {Fern{\'a}ndez-Garc{\'\i}a}, E.~J. and {Wu}, S. -Y. and {Almeida}, L. and {Aryan}, A. and {Belkin}, S. and {Bom}, C.~R. and {Butner}, M. and {Burkhonov}, O. and {Carrasco-Garc{\'\i}a}, I. and {Castell{\'o}n}, A. and {Castro Tirado}, M.~A. and {Chelovekov}, I. and {Egamberdiyev}, Sh. A. and {Garc{\'\i}a-Benito}, R. and {Garc{\'\i}a Gonz{\'a}lez}, S.~A. and {Grebenev}, S. and {Kilpatrick}, C.~D. and {Klunko}, E. and {Makler}, M. and {Minaev}, P. and {Mkrtchyan}, A. and {Moskvitin}, A. and {Navarete}, F. and {Novichonok}, A. and {Pankov}, N. and {Passas-Varo}, M. and {P{\'e}rez del Pulgar}, C. and {Reina Terol}, A. and {Smith}, J.~A. and {Tinyanont}, S. and {Tucker}, D.~L. and {Uklein}, R. and {Volnova}, A. and {Wiesner}, M.~P. and {Gritsevich}, M. and {Castro-Tirado}, A.~J.},
        title = "{Early photometric and spectroscopic observations of the extraordinarily bright INTEGRAL-detected GRB 221009A}",
      journal = {\aap},
         year = 2024,
        month = dec,
       volume = {692},
          eid = {A3},
        pages = {A3},
          doi = {10.1051/0004-6361/202449783},
       adsurl = {https://ui.adsabs.harvard.edu/abs/2024A&A...692A...3S}
}

@ARTICLE{sari1998,
       author = {{Sari}, Re'em and {Piran}, Tsvi and {Narayan}, Ramesh},
        title = "{Spectra and Light Curves of Gamma-Ray Burst Afterglows}",
      journal = {\apjl},
         year = 1998,
        month = apr,
       volume = {497},
       number = {1},
        pages = {L17-L20},
          doi = {10.1086/311269},
archivePrefix = {arXiv},
       eprint = {astro-ph/9712005},
 primaryClass = {astro-ph},
       adsurl = {https://ui.adsabs.harvard.edu/abs/1998ApJ...497L..17S}
}

@ARTICLE{sato2023,
       author = {{Sato}, Yuri and {Murase}, Kohta and {Ohira}, Yutaka and {Yamazaki}, Ryo},
        title = "{Two-component jet model for multiwavelength afterglow emission of the extremely energetic burst GRB 221009A}",
      journal = {\mnras},
         year = 2023,
        month = jun,
       volume = {522},
       number = {1},
        pages = {L56-L60},
          doi = {10.1093/mnrasl/slad038},
archivePrefix = {arXiv},
       eprint = {2212.09266},
 primaryClass = {astro-ph.HE},
       adsurl = {https://ui.adsabs.harvard.edu/abs/2023MNRAS.522L..56S}
}

@ARTICLE{sato2025,
       author = {{Sato}, Yuri and {Murase}, Kohta and {Ohira}, Yutaka and {Inoue}, Susumu and {Yamazaki}, Ryo},
        title = "{Two-component jet model for the afterglow emission of GRB 201216C and GRB 221009A and implications for jet structure of very-high-energy gamma-ray bursts}",
      journal = {Journal of High Energy Astrophysics},
         year = 2025,
        month = aug,
       volume = {48},
          eid = {100415},
        pages = {100415},
          doi = {10.1016/j.jheap.2025.100415},
archivePrefix = {arXiv},
       eprint = {2502.19051},
 primaryClass = {astro-ph.HE},
       adsurl = {https://ui.adsabs.harvard.edu/abs/2025JHEAp..4800415S}
}

@ARTICLEschlafly{schlafly2011,
       author = {{Schlafly}, Edward F. and {Finkbeiner}, Douglas P.},
        title = "{Measuring Reddening with Sloan Digital Sky Survey Stellar Spectra and Recalibrating SFD}",
      journal = {\apj},
         year = 2011,
        month = aug,
       volume = {737},
       number = {2},
          eid = {103},
        pages = {103},
          doi = {10.1088/0004-637X/737/2/103},
archivePrefix = {arXiv},
       eprint = {1012.4804},
 primaryClass = {astro-ph.GA},
       adsurl = {https://ui.adsabs.harvard.edu/abs/2011ApJ...737..103S}
}

@ARTICLE{sears2025,
       author = {{Sears}, Huei and {Chornock}, Ryan and {Blanchard}, Peter K. and {Margutti}, Raffaella and {Villar}, V. Ashley and {Pierel}, Justin and {Vallely}, Patrick J. and {Alexander}, Kate D. and {Berger}, Edo and {Eftekhari}, Tarraneh and {Jacobson-Gal{\'a}n}, Wynn V. and {Laskar}, Tanmoy and {LeBaron}, Natalie and {Metzger}, Brian D. and {Milisavljevic}, Dan},
        title = "{Late-time HST and JWST Observations of GRB 221009A: Evidence for a Break in the Light Curve at 50 days}",
      journal = {\apj},
         year = 2025,
        month = may,
       volume = {984},
       number = {2},
          eid = {196},
        pages = {196},
          doi = {10.3847/1538-4357/adc306},
archivePrefix = {arXiv},
       eprint = {2412.02663},
 primaryClass = {astro-ph.HE},
       adsurl = {https://ui.adsabs.harvard.edu/abs/2025ApJ...984..196S}
}

@ARTICLE{shrestha2023,
       author = {{Shrestha}, Manisha and {Sand}, David J. and {Alexander}, Kate D. and {Bostroem}, K. Azalee and {Hosseinzadeh}, Griffin and {Pearson}, Jeniveve and {Aghakhanloo}, Mojgan and {Vink{\'o}}, J{\'o}zsef and {Andrews}, Jennifer E. and {Jencson}, Jacob E. and {Lundquist}, M.~J. and {Wyatt}, Samuel and {Howell}, D. Andrew and {McCully}, Curtis and {Gonzalez}, Estefania Padilla and {Pellegrino}, Craig and {Terreran}, Giacomo and {Hiramatsu}, Daichi and {Newsome}, Megan and {Farah}, Joseph and {Jha}, Saurabh W. and {Smith}, Nathan and {Wheeler}, J. Craig and {Mart{\'\i}nez-V{\'a}zquez}, Clara and {Carballo-Bello}, Julio A. and {Drlica-Wagner}, Alex and {James}, David J. and {Mutlu-Pakdil}, Bur{\c{c}}in and {Stringfellow}, Guy S. and {Sakowska}, Joanna D. and {No{\"e}l}, Noelia E.~D. and {Bom}, Cl{\'e}cio R. and {Kuehn}, Kyler},
        title = "{Limit on Supernova Emission in the Brightest Gamma-Ray Burst, GRB 221009A}",
      journal = {\apjl},
         year = 2023,
        month = mar,
       volume = {946},
       number = {1},
          eid = {L25},
        pages = {L25},
          doi = {10.3847/2041-8213/acbd50},
archivePrefix = {arXiv},
       eprint = {2302.03829},
 primaryClass = {astro-ph.HE},
       adsurl = {https://ui.adsabs.harvard.edu/abs/2023ApJ...946L..25S}
}

@ARTICLE{siegel2019,
       author = {{Siegel}, Daniel M. and {Barnes}, Jennifer and {Metzger}, Brian D.},
        title = "{Collapsars as a major source of r-process elements}",
      journal = {\nat},
         year = 2019,
        month = may,
       volume = {569},
       number = {7755},
        pages = {241-244},
          doi = {10.1038/s41586-019-1136-0},
archivePrefix = {arXiv},
       eprint = {1810.00098},
 primaryClass = {astro-ph.HE},
       adsurl = {https://ui.adsabs.harvard.edu/abs/2019Natur.569..241S}
}

@ARTICLE{sneppen2025,
       author = {{Sneppen}, Albert and {Watson}, Darach},
        title = "{The optical constants and grain sizes of interstellar dust measured directly using the dust-scattered X-ray halo of GRB221009A}",
      journal = {\aap},
         year = 2025,
        month = sep,
       volume = {701},
          eid = {A65},
        pages = {A65},
          doi = {10.1051/0004-6361/202554741},
archivePrefix = {arXiv},
       eprint = {2506.12125},
 primaryClass = {astro-ph.HE},
       adsurl = {https://ui.adsabs.harvard.edu/abs/2025A&A...701A..65S}
}

@ARTICLE{sironi2011,
       author = {{Sironi}, Lorenzo and {Spitkovsky}, Anatoly},
        title = "{Particle Acceleration in Relativistic Magnetized Collisionless Electron-Ion Shocks}",
      journal = {\apj},
         year = 2011,
        month = jan,
       volume = {726},
       number = {2},
          eid = {75},
        pages = {75},
          doi = {10.1088/0004-637X/726/2/75},
archivePrefix = {arXiv},
       eprint = {1009.0024},
 primaryClass = {astro-ph.HE},
       adsurl = {https://ui.adsabs.harvard.edu/abs/2011ApJ...726...75S}
}

@ARTICLE{srinivasaragavan2023,
       author = {{Srinivasaragavan}, Gokul P. and {O'Connor}, Brendan and {Cenko}, S. Bradley and {Dittmann}, Alexander J. and {Yang}, Sheng and {Sollerman}, Jesper and {Anupama}, G.~C. and {Barway}, Sudhanshu and {Bhalerao}, Varun and {Kumar}, Harsh and {Swain}, Vishwajeet and {Hammerstein}, Erica and {Holt}, Isiah and {Anand}, Shreya and {Andreoni}, Igor and {Coughlin}, Michael W. and {Dichiara}, Simone and {Gal-Yam}, Avishay and {Miller}, M. Coleman and {Soon}, Jaime and {Soria}, Roberto and {Durbak}, Joseph and {Gillanders}, James H. and {Laha}, Sibasish and {Moore}, Anna M. and {Ragosta}, Fabio and {Troja}, Eleonora},
        title = "{A Sensitive Search for Supernova Emission Associated with the Extremely Energetic and Nearby GRB 221009A}",
      journal = {\apjl},
         year = 2023,
        month = jun,
       volume = {949},
       number = {2},
          eid = {L39},
        pages = {L39},
          doi = {10.3847/2041-8213/accf97},
archivePrefix = {arXiv},
       eprint = {2303.12849},
 primaryClass = {astro-ph.HE},
       adsurl = {https://ui.adsabs.harvard.edu/abs/2023ApJ...949L..39S}
}

@ARTICLE{tiengo2023,
       author = {{Tiengo}, Andrea and {Pintore}, Fabio and {Vaia}, Beatrice and {Filippi}, Simone and {Sacchi}, Andrea and {Esposito}, Paolo and {Rigoselli}, Michela and {Mereghetti}, Sandro and {Salvaterra}, Ruben and {{\v{S}}iljeg}, Barbara and {Bracco}, Andrea and {Bo{\v{s}}njak}, {\v{Z}}eljka and {Jeli{\'c}}, Vibor and {Campana}, Sergio},
        title = "{The Power of the Rings: The GRB 221009A Soft X-Ray Emission from Its Dust-scattering Halo}",
      journal = {\apjl},
         year = 2023,
        month = mar,
       volume = {946},
       number = {1},
          eid = {L30},
        pages = {L30},
          doi = {10.3847/2041-8213/acc1dc},
archivePrefix = {arXiv},
       eprint = {2302.11518},
 primaryClass = {astro-ph.HE},
       adsurl = {https://ui.adsabs.harvard.edu/abs/2023ApJ...946L..30T}
}

@ARTICLE{vaia2025,
       author = {{Vaia}, B. and {Bo{\v{s}}njak}, {\v{Z}}. and {Bracco}, A. and {Campana}, S. and {Esposito}, P. and {Jeli{\'c}}, V. and {Sacchi}, A. and {Tiengo}, A.},
        title = "{Probing the interstellar medium toward GRB 221009A through X-ray dust scattering}",
      journal = {\aap},
         year = 2025,
        month = apr,
       volume = {696},
          eid = {A9},
        pages = {A9},
          doi = {10.1051/0004-6361/202453158},
archivePrefix = {arXiv},
       eprint = {2502.20940},
 primaryClass = {astro-ph.HE},
       adsurl = {https://ui.adsabs.harvard.edu/abs/2025A&A...696A...9V}
}

@ARTICLE{vaneerten2013,
       author = {{van Eerten}, Hendrik and {MacFadyen}, Andrew},
        title = "{Gamma-Ray Burst Afterglow Light Curves from a Lorentz-boosted Simulation Frame and the Shape of the Jet Break}",
      journal = {\apj},
         year = 2013,
        month = apr,
       volume = {767},
       number = {2},
          eid = {141},
        pages = {141},
          doi = {10.1088/0004-637X/767/2/141},
archivePrefix = {arXiv},
       eprint = {1209.1985},
 primaryClass = {astro-ph.HE},
       adsurl = {https://ui.adsabs.harvard.edu/abs/2013ApJ...767..141V}
}

@ARTICLE{vasilopoulos2023,
       author = {{Vasilopoulos}, Georgios and {Karavola}, Despina and {Stathopoulos}, Stamatios I. and {Petropoulou}, Maria},
        title = "{Dust-scattering rings of GRB 221009A as seen by the Neil Gehrels Swift X-ray Observatory: can we count them all?}",
      journal = {\mnras},
         year = 2023,
        month = may,
       volume = {521},
       number = {1},
        pages = {1590-1600},
          doi = {10.1093/mnras/stad375},
archivePrefix = {arXiv},
       eprint = {2302.02383},
 primaryClass = {astro-ph.HE},
       adsurl = {https://ui.adsabs.harvard.edu/abs/2023MNRAS.521.1590V}
}

@article{williams2021,
    author = "Williams, Michael J. and Veitch, John and Messenger, Chris",
    title = "{Nested sampling with normalizing flows for gravitational-wave inference}",
    eprint = "2102.11056",
    archivePrefix = "arXiv",
    primaryClass = "gr-qc",
    doi = "10.1103/PhysRevD.103.103006",
    journal = "Phys. Rev. D",
    volume = "103",
    number = "10",
    pages = "103006",
    year = "2021"
}

@article{williams2023,
    author = {{Williams}, Michael J. and {Veitch}, John and {Messenger}, Chris},
    title = "{Importance nested sampling with normalising flows}",
    journal = {Machine Learning: Science and Technology},
    year = 2023,
    month = sep,
    volume = {4},
    number = {3},
    eid = {035011},
    pages = {035011},
    doi = {10.1088/2632-2153/acd5aa},
    archivePrefix = {arXiv},
    eprint = {2302.08526},
    primaryClass = {astro-ph.IM},
    adsurl = {https://ui.adsabs.harvard.edu/abs/2023MLS&T...4c5011W}
}

@ARTICLE{williams_maia2023,
       author = {{Williams}, Maia A. and {Kennea}, Jamie A. and {Dichiara}, S. and {Kobayashi}, Kohei and {Iwakiri}, Wataru B. and {Beardmore}, Andrew P. and {Evans}, P.~A. and {Heinz}, Sebastian and {Lien}, Amy and {Oates}, S.~R. and {Negoro}, Hitoshi and {Cenko}, S. Bradley and {Buisson}, Douglas J.~K. and {Hartmann}, Dieter H. and {Jaisawal}, Gaurava K. and {Kuin}, N.~P.~M. and {Lesage}, Stephen and {Page}, Kim L. and {Parsotan}, Tyler and {Pasham}, Dheeraj R. and {Sbarufatti}, B. and {Siegel}, Michael H. and {Sugita}, Satoshi and {Younes}, George and {Ambrosi}, Elena and {Arzoumanian}, Zaven and {Bernardini}, M.~G. and {Campana}, S. and {Capalbi}, Milvia and {Caputo}, Regina and {D'A{\`\i}}, Antonino and {D'Avanzo}, P. and {D'Elia}, V. and {De Pasquale}, Massimiliano and {Eyles-Ferris}, R.~A.~J. and {Ferrara}, Elizabeth and {Gendreau}, Keith C. and {Gropp}, Jeffrey D. and {Kawai}, Nobuyuki and {Klingler}, Noel and {Laha}, Sibasish and {Melandri}, A. and {Mihara}, Tatehiro and {Moss}, Michael and {O'Brien}, Paul and {Osborne}, Julian P. and {Palmer}, David M. and {Perri}, Matteo and {Serino}, Motoko and {Sonbas}, E. and {Stamatikos}, Michael and {Starling}, Rhaana and {Tagliaferri}, G. and {Tohuvavohu}, Aaron and {Zane}, Silvia and {Ziaeepour}, Houri},
        title = "{GRB 221009A: Discovery of an Exceptionally Rare Nearby and Energetic Gamma-Ray Burst}",
      journal = {\apjl},
         year = 2023,
        month = mar,
       volume = {946},
       number = {1},
          eid = {L24},
        pages = {L24},
          doi = {10.3847/2041-8213/acbcd1},
archivePrefix = {arXiv},
       eprint = {2302.03642},
 primaryClass = {astro-ph.HE},
       adsurl = {https://ui.adsabs.harvard.edu/abs/2023ApJ...946L..24W}
}

@ARTICLE{zhang2024,
       author = {{Zhang}, Bing and {Wang}, Xiang-Yu and {Zheng}, Jian-He},
        title = "{The BOAT GRB 221009A: A Poynting-flux-dominated narrow jet surrounded by a matter-dominated structured jet wing}",
      journal = {Journal of High Energy Astrophysics},
         year = 2024,
        month = mar,
       volume = {41},
        pages = {42-53},
          doi = {10.1016/j.jheap.2024.01.002},
archivePrefix = {arXiv},
       eprint = {2311.14180},
 primaryClass = {astro-ph.HE},
       adsurl = {https://ui.adsabs.harvard.edu/abs/2024JHEAp..41...42Z}
}

@ARTICLE{zhang2023_hai-ming,
       author = {{Zhang}, Hai-Ming and {Huang}, Yi-Yun and {Liu}, Ruo-Yu and {Wang}, Xiang-Yu},
        title = "{GRB 221009A: Revealing a Hidden Afterglow during the Prompt Emission Phase with Fermi-GBM Observations}",
      journal = {\apjl},
         year = 2023,
        month = oct,
       volume = {956},
       number = {1},
          eid = {L21},
        pages = {L21},
          doi = {10.3847/2041-8213/acfcab},
archivePrefix = {arXiv},
       eprint = {2307.12623},
 primaryClass = {astro-ph.HE},
       adsurl = {https://ui.adsabs.harvard.edu/abs/2023ApJ...956L..21Z}
}

@ARTICLE{zhao2024,
       author = {{Zhao}, Guoying and {Shen}, Rong-Feng},
        title = "{Expansion and Spectral Softening of the Dust-scattering Rings of GRB 221009A}",
      journal = {\apj},
         year = 2024,
        month = aug,
       volume = {970},
       number = {2},
          eid = {124},
        pages = {124},
          doi = {10.3847/1538-4357/ad5a85},
archivePrefix = {arXiv},
       eprint = {2403.09444},
 primaryClass = {astro-ph.HE},
       adsurl = {https://ui.adsabs.harvard.edu/abs/2024ApJ...970..124Z}
}

@ARTICLE{zheng2024,
       author = {{Zheng}, Jian-He and {Wang}, Xiang-Yu and {Liu}, Ruo-Yu and {Zhang}, Bing},
        title = "{A Narrow Uniform Core with a Wide Structured Wing: Modeling the TeV and Multiwavelength Afterglows of GRB 221009A}",
      journal = {\apj},
         year = 2024,
        month = may,
       volume = {966},
       number = {1},
          eid = {141},
        pages = {141},
          doi = {10.3847/1538-4357/ad3949},
archivePrefix = {arXiv},
       eprint = {2310.12856},
 primaryClass = {astro-ph.HE},
       adsurl = {https://ui.adsabs.harvard.edu/abs/2024ApJ...966..141Z}
}

@ARTICLE{zheng2024_chao,
       author = {{Zheng}, Chao and {Zhang}, Yan-Qiu and {Xiong}, Shao-Lin and {Li}, Cheng-Kui and {Gao}, He and {Xue}, Wang-Chen and {Liu}, Jia-Cong and {Wang}, Chen-Wei and {Tan}, Wen-Jun and {Peng}, Wen-Xi and {An}, Zheng-Hua and {Cai}, Ce and {Ge}, Ming-Yu and {Guo}, Dong-Ya and {Huang}, Yue and {Li}, Bing and {Li}, Ti-Pei and {Li}, Xiao-Bo and {Li}, Xin-Qiao and {Li}, Xu-Fang and {Liao}, Jin-Yuan and {Liu}, Cong-Zhan and {Lu}, Fang-Jun and {Ma}, Xiang and {Qiao}, Rui and {Song}, Li-Ming and {Wang}, Jin and {Wang}, Ping and {Wang}, Xi-Lu and {Wang}, Yue and {Wen}, Xiang-Yang and {Xiao}, Shuo and {Xu}, Yan-Bing and {Xu}, Yu-Peng and {Yao}, Zhi-Guo and {Yi}, Qi-Bing and {Yi}, Shu-Xu and {You}, Yuan and {Zhang}, Fan and {Zhang}, Jin-Peng and {Zhang}, Peng and {Zhang}, Shu and {Zhang}, Shuang-Nan and {Zhang}, Yan-Ting and {Zhang}, Zhen and {Zhao}, Xiao-Yun and {Zhao}, Yi and {Zheng}, Shi-Jie},
        title = "{Observation of GRB 221009A Early Afterglow in X-Ray/Gamma-Ray Energy Bands}",
      journal = {\apjl},
         year = 2024,
        month = feb,
       volume = {962},
       number = {1},
          eid = {L2},
        pages = {L2},
          doi = {10.3847/2041-8213/ad2073},
archivePrefix = {arXiv},
       eprint = {2310.10522},
 primaryClass = {astro-ph.HE},
       adsurl = {https://ui.adsabs.harvard.edu/abs/2024ApJ...962L...2Z}
}

@article{Burns_2023,
   title={GRB 221009A: The BOAT},
   volume={946},
   ISSN={2041-8213},
   url={http://dx.doi.org/10.3847/2041-8213/acc39c},
   DOI={10.3847/2041-8213/acc39c},
   number={1},
   journal={The Astrophysical Journal Letters},
   publisher={American Astronomical Society},
   author={Burns, Eric and Svinkin, Dmitry and Fenimore, Edward and Kann, D. Alexander and Agüí Fernández, José Feliciano and Frederiks, Dmitry and Hamburg, Rachel and Lesage, Stephen and Temiraev, Yuri and Tsvetkova, Anastasia and Bissaldi, Elisabetta and Briggs, Michael S. and Dalessi, Sarah and Dunwoody, Rachel and Fletcher, Cori and Goldstein, Adam and Hui, C. Michelle and Hristov, Boyan A. and Kocevski, Daniel and Lysenko, Alexandra L. and Mailyan, Bagrat and Mangan, Joseph and McBreen, Sheila and Racusin, Judith and Ridnaia, Anna and Roberts, Oliver J. and Ulanov, Mikhail and Veres, Peter and Wilson-Hodge, Colleen A. and Wood, Joshua},
   year={2023},
   month=mar, pages={L31} }

@article{Hayes_2022,
doi = {10.3847/2515-5172/ac9d2f},
url = {https://doi.org/10.3847/2515-5172/ac9d2f},
year = {2022},
month = {oct},
publisher = {The American Astronomical Society},
volume = {6},
number = {10},
pages = {222},
author = {Hayes, Laura A. and Gallagher, Peter T.},
title = {A Significant Sudden Ionospheric Disturbance Associated with Gamma-Ray Burst GRB 221009A},
journal = {Research Notes of the AAS}
}

@article{Frederiks_2023,
doi = {10.3847/2041-8213/acd1eb},
url = {https://doi.org/10.3847/2041-8213/acd1eb},
year = {2023},
month = {may},
publisher = {The American Astronomical Society},
volume = {949},
number = {1},
pages = {L7},
author = {Frederiks, D. and Svinkin, D. and Lysenko, A. L. and Molkov, S. and Tsvetkova, A. and Ulanov, M. and Ridnaia, A. and Lutovinov, A. A. and Lapshov, I. and Tkachenko, A. and Levin, V.},
title = {Properties of the Extremely Energetic GRB 221009A from Konus-WIND and SRG/ART-XC Observations},
journal = {The Astrophysical Journal Letters}
}

@ARTICLE{Campana_2024,
       author = {{Campana}, Sergio and {Braito}, Valentina and {Lazzati}, Davide and {Tiengo}, Andrea},
        title = "{A Search for Soft X-Ray Emission Lines in the Afterglow Spectrum of GRB 221009A}",
      journal = {\apj},
         year = 2024,
        month = sep,
       volume = {972},
       number = {1},
          eid = {75},
        pages = {75},
          doi = {10.3847/1538-4357/ad6b96},
archivePrefix = {arXiv},
       eprint = {2408.03306},
 primaryClass = {astro-ph.HE},
       adsurl = {https://ui.adsabs.harvard.edu/abs/2024ApJ...972...75C}
}

@ARTICLE{Paciesas_1999,
       author = {{Paciesas}, William S. and {Meegan}, Charles A. and {Pendleton}, Geoffrey N. and {Briggs}, Michael S. and {Kouveliotou}, Chryssa and {Koshut}, Thomas M. and {Lestrade}, John Patrick and {McCollough}, Michael L. and {Brainerd}, Jerome J. and {Hakkila}, Jon and {Henze}, William and {Preece}, Robert D. and {Connaughton}, Valerie and {Kippen}, R. Marc and {Mallozzi}, Robert S. and {Fishman}, Gerald J. and {Richardson}, Georgia A. and {Sahi}, Maitrayee},
        title = "{The Fourth BATSE Gamma-Ray Burst Catalog (Revised)}",
      journal = {\apjs},
         year = 1999,
        month = jun,
       volume = {122},
       number = {2},
        pages = {465-495},
          doi = {10.1086/313224},
archivePrefix = {arXiv},
       eprint = {astro-ph/9903205},
 primaryClass = {astro-ph},
       adsurl = {https://ui.adsabs.harvard.edu/abs/1999ApJS..122..465P}
}

@ARTICLE{Paciesas_2012,
       author = {{Paciesas}, William S. and {Meegan}, Charles A. and {von Kienlin}, Andreas and {Bhat}, P.~N. and {Bissaldi}, Elisabetta and {Briggs}, Michael S. and {Burgess}, J. Michael and {Chaplin}, Vandiver and {Connaughton}, Valerie and {Diehl}, Roland and {Fishman}, Gerald J. and {Fitzpatrick}, Gerard and {Foley}, Suzanne and {Gibby}, Melissa and {Giles}, Misty and {Goldstein}, Adam and {Greiner}, Jochen and {Gruber}, David and {Guiriec}, Sylvain and {van der Horst}, Alexander J. and {Kippen}, R. Marc and {Kouveliotou}, Chryssa and {Lichti}, Giselher and {Lin}, Lin and {McBreen}, Sheila and {Preece}, Robert D. and {Rau}, Arne and {Tierney}, Dave and {Wilson-Hodge}, Colleen},
        title = "{The Fermi GBM Gamma-Ray Burst Catalog: The First Two Years}",
      journal = {\apjs},
         year = 2012,
        month = mar,
       volume = {199},
       number = {1},
          eid = {18},
        pages = {18},
          doi = {10.1088/0067-0049/199/1/18},
archivePrefix = {arXiv},
       eprint = {1201.3099},
 primaryClass = {astro-ph.HE},
       adsurl = {https://ui.adsabs.harvard.edu/abs/2012ApJS..199...18P}
}

@ARTICLE{Hoffman_Draine_2016,
       author = {{Hoffman}, John and {Draine}, B.~T.},
        title = "{Accurate Modeling of X-ray Extinction by Interstellar Grains}",
      journal = {\apj},
         year = 2016,
        month = feb,
       volume = {817},
       number = {2},
          eid = {139},
        pages = {139},
          doi = {10.3847/0004-637X/817/2/139},
archivePrefix = {arXiv},
       eprint = {1509.08987},
 primaryClass = {astro-ph.HE},
       adsurl = {https://ui.adsabs.harvard.edu/abs/2016ApJ...817..139H}
}

@ARTICLE{Costantini_2019,
       author = {{Costantini}, E. and {Zeegers}, S.~T. and {Rogantini}, D. and {de Vries}, C.~P. and {Tielens}, A.~G.~G.~M. and {Waters}, L.~B.~F.~M.},
        title = "{X-ray extinction from interstellar dust. Prospects of observing carbon, sulfur, and other trace elements}",
      journal = {\aap},
         year = 2019,
        month = sep,
       volume = {629},
          eid = {A78},
        pages = {A78},
          doi = {10.1051/0004-6361/201833820},
archivePrefix = {arXiv},
       eprint = {1906.08653},
 primaryClass = {astro-ph.GA},
       adsurl = {https://ui.adsabs.harvard.edu/abs/2019A&A...629A..78C}
}




\appendix

\section{X-Shooter epoch i, ii, and iii, and NIRSPEC + MIRI at the first JWST epoch}

\begin{figure*}
    \centering
    \includegraphics[width=\linewidth]{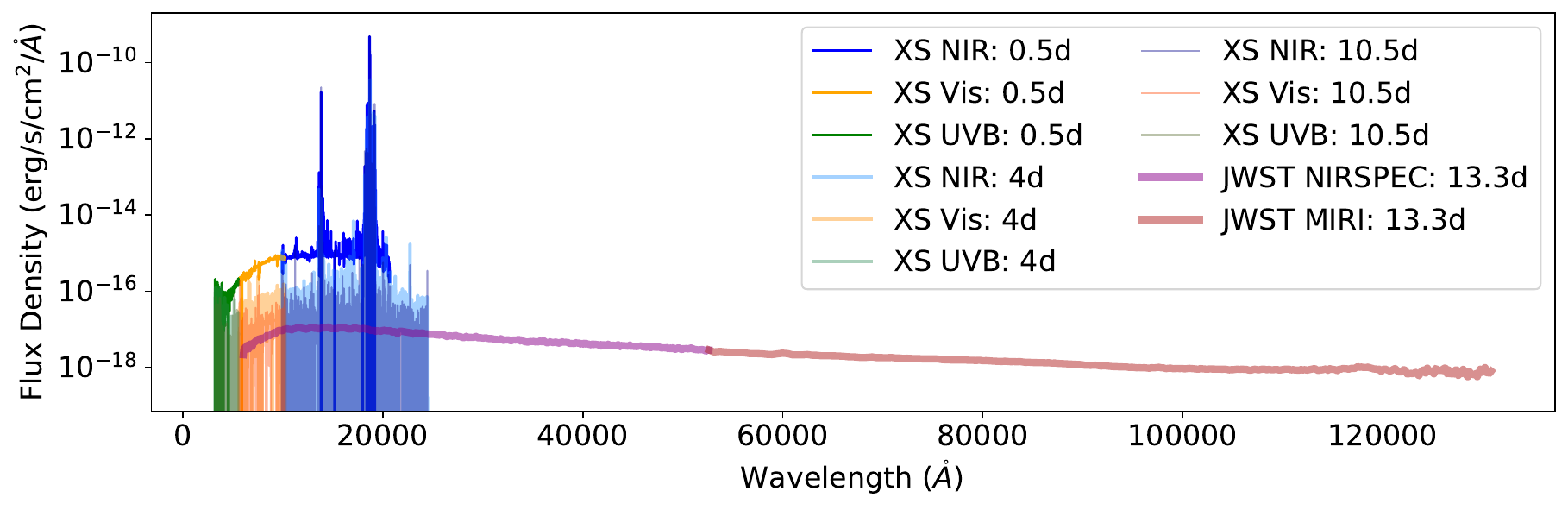}
    \caption{Afterglow spectral data from the three channels of X-Shooter; NIR, visible, and UV-B and the JWST NIRSPEC and MIRI data at 13.3 days.}
    \label{fig:XS123}
\end{figure*}

A joint plot of the optical to mid-infrared afterglow spectra from X-Shooter and JWST is shown in Figure~\ref{fig:XS123}. It serves to illustrate the stark difference in the SNR obtained by JWST compared to X-Shooter.

\section{Individual epoch fits}
The individual epoch fits for the X-Shooter and JWST spectra are used together to find viable scenarios, given what we know and what we expect for GRB afterglow evolution.
To find the best fitting cases for each of the epochs and each of the models, PL (Eq.\ref{eq:simp}), SBPL1 (Eq.\ref{eq:low}), and SBPL2 (Eq.\ref{eq:high}), we plot the normalised log of the Bayesian evidence from the {\sc Nessai} results and show the comparison at fixed extinction values for each epoch in Figure~\ref{fig:log_e}.
We normalise the log evidence values by the maximum log evidence -- where the closer to unity that the value is, the stronger the fit.

The three model fit results are presented visually in Figure\,\ref{fig:AV_xb} for the fixed extinction, $A_V$ vs the median spectral index from the posterior distribution, $\beta$.
Each row corresponds to a model; PL, SBPL1, SBPL2.
The hatched regions represent the parameter space outside of the typical range for the spectral index of an afterglow for the observed GRB population, leaving clear the range $0.2<\beta<1.2$ \citep{kann2010}.
Two of the typical spectral index values inferred from afterglow modelling are shown as dotted and dash-dotted horizontal lines at $\beta=0.76$ \citep[e.g.,][]{laskar2023}, and $\beta=0.36$ \citep[e.g.,][]{levan2023}.
For each epoch, the $\beta$ values are shown with a different line style; solid line for epoch i; thick dash-dotted line for epoch ii; thick dotted line for epoch iii; and a thick dashed line for the 13 day JWST data.
The vertical coloured panels are the published ranges for $A_V$ from \citet{levan2023} -- pink; \citet{kann2023} -- yellow; and \citet{blanchard2024} -- teal.

In Figure\,\ref{fig:AV_xb}, the PL model (top panel) clearly shows that for all the X-Shooter epochs, the inferred spectral index is a function of the assumed extinction, $\beta(A_V)$.
This dependence is reduced, although still marginally present, for the JWST data, which starts at redder wavelengths and extends to much longer wavelengths than the X-Shooter data.
For X-Shooter data the range of inferred spectral index is, $\Delta\beta \simeq 1.4$, while for JWST, the range is $\Delta\beta\simeq0.4$, over the same extinction range, $3.5\leq A_V \leq 7.0$. Thus, the X-Shooter bands have a stronger degeneracy between the extinction and the spectral index.
For the SBPL1 and SBPL2 cases, this degeneracy is not as clear due to the break in the model and the smoothness of the transition between the two indices.
For all models, the three X-Shooter epochs return distinct solutions, while the variance in the JWST fit for all the models is largely similar from $A_V\gtrsim4.5$, with the preferred spectral index at lower values of extinction becoming increasingly sensitive to the break transition or the second spectral index. 

The X-Shooter data at epoch i and iii were originally modelled by \citet{malesani2025} with an assumed spectral index of $\beta\sim0.8$ for an $A_V = 4.177$ and $\sim 0.4$ for $A_V = 4.7$. From our PL fits, for $A_V=4.177$ we find that epoch iii returns $\beta_{\rm{iii}}\sim0.8$, however, epoch i and ii both show smaller spectral indices, $\beta_{\rm{i}} \sim0.35$ and $\beta_{\rm{ii}}\sim0.65$; while for $A_V = 4.7$, the model fits return $\beta_{\rm{i}}\sim0.15$, $\beta_{\rm{ii}}\sim0.45$, and $\beta_{\rm{iii}}\sim0.7$ vs the $\beta=0.4$ used by \citet{malesani2025} (who did not fit their values). A fit was performed by \citet{levan2023} on the X-Shooter data that confirmed the higher extinction values, returning $A_V=4.9$ and $\beta_{\rm{i}}=0.21$. At this extinction, our revised fit using a nested sampling algorithm instead of a Markov-Chain Monte Carlo returns $\beta_{\rm{i}}\sim0.1$, however, for epoch i the two values of $R_V\sim3.2$ are consistent.

The JWST data shown in our figures is that from the new pipeline reduction mentioned in Section \ref{s:JWST_fiiting}, this data shows an offset in $A_V - \beta$ space of $\Delta\beta \sim 0.05$ when compared to spectral indices found by fitting the older version of the data reduction.
For the original JWST data, the PL model returns spectral indices for a given extinction that are consistent with both the results of \citet{levan2023} at $\beta=0.36$ for an $A_V=4.94$, and \citet{blanchard2024} with $\beta=0.41$ for an $A_V=4.63$, while the offset for the new pipeline data is seen at all extinction values and returns $\beta \sim 0.41$ at $A_V= 4.9$, and $\beta\sim0.46$ for $A_V = 4.6$, respectively.
The difference between the \citet{levan2023} and \citet{blanchard2024} results (model fit $\beta$ and $A_V$) are likely routed in the assumptions for the prior ranges made by either author for the reddening. \citet{blanchard2024} did not specify explicit prior ranges, but stated that they used a wide uniform prior for all of their parameters (except the normalisation, which is a log-uniform prior). Whereas \citet{levan2023} set the priors of the spectral index as $0.0 \leq \beta \leq 1.0$; the MW range of extinction as $0.0 \leq A_{V,MW} \leq 10.0$; and setting the fit condition such that $A_V \geq R_V E(B-V)$.
For the PL models, the X-Shooter epoch ii $\beta$ coincides with that of JWST at the extinction value found by \citet{levan2023}.

\begingroup
\begin{table*}
    \centering
    \caption{The power-law index with fixed $A_V$ for all epochs of X-Shooter and the 13 day JWST data.
    The first spectral index ($\beta$) column is for a single power law.
    The second $\beta_1$ column is for a smoothly broken power-law, where the long wavelength index is fixed to $\beta = -1/3$.
    The third column lists the break wavelength ($\mu$m) for the $(p-1)/2 \rightarrow -1/3$.
    The fourth $\beta_2$ column is for a smoothly broken power law where the short wavelength spectral index is $\beta = p/2$ and $p = 2\beta_2 + 1$.
    The fifth column has the break wavelength for the $p/2 \rightarrow (p-1)/2$ transition.
    Where only a single value is given, the level of uncertainty is below the precision of the listed parameter.
    Values in {\bf bold} indicate those within the physical expectation range from the literature, $0.20\lesssim\beta\lesssim0.90$ \citep{laskar2023, levan2023, shrestha2023, sanchez-ramirez2024}.
    }
    \label{tab:XSe1_fit}
    \renewcommand{\arraystretch}{1.32}
    \begin{tabular}{c|c|c|c|c|c|c}
        Instr. & $A_V$ & $\beta$ & $\beta_1$ & $\lambda_{\beta_1}$ & $\beta_2$ & $\lambda_{\beta_2}$ \\
        (epoch) & (mag)  & $\left[\lambda^\beta\right]$ & $\left[\lambda^{\beta_1} \rightarrow \lambda^{-1/3}\right]$   & ($\mu$m)  & $\left[\lambda^{(2\beta_2+1)/2} \rightarrow \lambda^{\beta_2}\right]$ &  ($\mu$m)  \\
        \hline
        X-Shooter (i) &  & $\bf 0.65\pm{0.01}$&  $0.99$& $4.02^{+0.02}_{-0.03}$& {$\bf 0.406\pm{0.00}$}& {$\bf1.21\pm0.07$}\\
        X-Shooter (ii) & 3.5& $ 0.94\pm0.01$&  $0.99$& $24.81^{+2.89}_{-2.37}$& {$\bf 0.68 \pm{0.03}$}& {$\bf 1.32^{+0.67}_{-0.44}$}\\
        X-Shooter (iii) & & {$ 1.05\pm0.02$}& $0.99$& $784.33^{+449.13}_{-375.69}$& {$\bf 0.79 \pm 0.07$}& {$\bf 1.58^{+2.85}_{-1.03}$}\\
        NIRSPEC + MIRI (i) &   & {$\bf 0.58$}& {$ \bf 0.78$}& {$\bf 17.11^{+0.18}_{-0.16}$}& {{$\bf 0.322 \pm{0.002}$}}& {{$\bf 3.23\pm{0.12}$}}\\
        \hline
        X-Shooter (i) &  & {$\bf 0.430^{+0.002}_{-0.001}$}&  $0.99$&  $1.85\pm{0.01}$& $0.183\pm{0.005}$& $1.26^{+0.10}_{-0.09}$\\
        X-Shooter (ii) & 4.0  & {$\bf 0.71\pm0.01$}& $0.99^{+0.01}_{-0.01}$& $5.79^{+0.33}_{-0.29}$& { $\bf 0.46^{+0.03}_{-0.04}$}&  {$\bf 1.42^{+0.95}_{-0.58}$}\\
        X-Shooter (iii) & &\ { {$\bf0.87\pm0.02$}}& $0.97^{+0.02}_{-0.04}$& $24.50 ^{+13.51}_{-5.36}$& { $\bf 0.57^{+0.13}_{-0.10}$}&  {$\bf 3.09^{+18.72}_{-2.71}$}\\
        NIRSPEC + MIRI (i) &   &{{$\bf 0.52$}}& {{$\bf 0.61$}}& {{$\bf 40.87^{+0.75}_{-0.73}$}}& {$\bf 0.22\pm0.01$}& { {$\bf 6.03^{+0.69}_{-0.64}$}}\\
        \hline
        X-Shooter (i) &   & {$\bf 0.230\pm{0.002}$}& {$ 0.991^{+0.006}_{-0.010}$}& $1.01\pm{0.01}$& $0.002^{+0.003}_{-0.001}$& $1.03^{+0.03}_{-0.04}$\\
        X-Shooter (ii) & 4.5 &  {$\bf 0.51\pm0.01$}& $0.97^{+0.02}_{-0.04}$& $2.90^{+0.28}_{-0.19}$&{ $\bf 0.25 \pm{0.05}$}& {$\bf 1.66^{+1.91}_{-0.89}$}\\
        X-Shooter (iii) & & { {$\bf0.72\pm0.02$}}&  {$\bf 0.86^{+0.09}_{-0.10}$}&  {$\bf 15.84^{+49.63}_{-6.18}$}&{ $\bf 0.36^{+0.27}_{-0.10}$}& { $\bf 11.67^{+199.63}_{-11.58}$}\\
        NIRSPEC + MIRI (i) &   &{$\bf 0.46$}& { {$\bf 0.46$}}& { {$\bf 1442.72^{+42.52}_{-93.33}$}}& {{$\bf 0.45$}}& {{$\bf 0.001$}}\\
        \hline
        X-Shooter (i) &  & $0.060\pm{0.002}$& {$\bf 0.50\pm0.02$}& {$\bf 1.34\pm{0.08}$}& $0.0008\pm{0.001} $& $0.042\pm{0.003}$\\
        X-Shooter (ii) & 5.0  & {$\bf 0.34\pm0.01$}& { {$\bf 0.68^{+0.11}_{-0.09}$}}& {$\bf 3.32^{+1.46}_{-0.87}$}& $0.08^{+0.11}_{-0.06}$&  $1.64^{+2.29}_{-1.34}$\\
        X-Shooter (iii) & & {{$\bf0.61\pm0.02$}}& {{$\bf0.69^{+0.12}_{-0.07}$}}& {$\bf 25.26^{+226.49}_{-16.23}$}&{ $\bf 0.18^{+0.21}_{-0.05}$}&{ $\bf 41.68^{+396.14}_{-40.89}$}\\
        NIRSPEC + MIRI (i) &   &  {$\bf 0.40$}& { {$\bf 0.41$}}& { {$\bf 1495.66^{+3.23}_{-7.34}$}}& {$\bf 0.40$}& {{$\bf 0.001$}}\\
        \hline
        X-Shooter (i) &  & $-0.083\pm{0.002}$&  $0.002^{+0.003}_{-0.002}$&  $14.88^{+1.40}_{-1.60}$& $(3.1^{+4.9}_{-2.3}) \times 10^{-5}$&  $0.001$\\
        X-Shooter (ii) & 5.5  & $ 0.19\pm0.01$& $\bf 0.24^{+0.08}_{-0.04}$& $\bf 48.01^{+379.62}_{-38.34}$& $0.17^{+0.02}_{-0.03}$& $0.005^{+0.032}_{-0.003}$\\
        X-Shooter (iii) & & $\bf 0.38\pm0.01$& $\bf 0.39 \pm 0.01$& $\bf 461.97^{+618.85}_{-308.88}$& $\bf 0.37 \pm 0.01$& $\bf 0.003^{+0.005}_{-0.001}$\\
        NIRSPEC + MIRI (i) &   & $\bf 0.34$&  $\bf 0.35$&  $\bf 1498.06^{+1.48}_{-3.09}$& $\bf 0.34$&  $\bf 0.001$\\
        \hline
        X-Shooter (i) &  & $-0.209\pm{0.002}$& $0.001^{+0.002}_{-0.001}$& $0.53\pm{0.03}$& $(4.9^{+7.1}_{-3.5}) \times 10^{-6}$& $0.001$\\
        X-Shooter (ii) & 6.0  & $0.08\pm0.01$& $ 0.13^{+0.05}_{-0.03}$& $59.40^{+137.84}_{-45.32}$&   $0.06 \pm{0.02}$& $0.003^{+0.010}_{-0.002}$\\
        X-Shooter (iii) & & $0.11\pm0.01$& $ 0.13 \pm{0.01}$& $721.14^{+487.17}_{-416.35}$&  $0.10 \pm{0.01}$& $0.002^{+0.002}_{-0.001}$\\
        NIRSPEC + MIRI (i)  &  &  $\bf 0.29$& $\bf 0.29$& $\bf 1498.79^{+0.90}_{-2.04}$&  {$\bf 0.28$}& $\bf 0.001$\\
        \hline
        X-Shooter (i) &  & $-0.467\pm{0.001}$&  $\bf 0.64\pm{0.02}$& $\bf 0.001$& $(4.6^{+4.2}_{-2.8}) \times 10^{-6}$& $0.001$\\
        X-Shooter (ii) & 6.5  &  $-0.15\pm0.01$&  $0.01^{+0.02}_{-0.01}$& $1.89^{+0.46}_{-0.43}$& $0.0002^{+0.0004}_{-0.0002}$&  $0.001$\\
        X-Shooter (iii) & & $-0.15\pm0.01$& $0.01 \pm 0.01$& $1.68^{+0.58}_{-0.43}$& $0.0005^{+0.0008}_{-0.0004}$& $0.001$\\
        NIRSPEC + MIRI (i) &   &  $\bf 0.24$& $\bf 0.24$& $\bf 1499.06^{+0.69}_{-1.50}$& $\bf 0.23$& $\bf 0.001$\\
        \hline
        X-Shooter (i) &  & $-0.735\pm{0.001}$& $\bf0.64 \pm{0.01}$& $\bf 0.001$&  $(1.5^{+2.6}_{-1.1}) \times 10^{-6}$& $0.001$\\
        X-Shooter (ii) & 7.0& $-0.41\pm0.01$&  $\bf0.63\pm{0.15}$& $\bf 0.001$& $0.0003^{+0.0003}_{-0.0002}$& $0.001$\\
        X-Shooter (iii) & & $-0.42\pm0.01$& $\bf0.63 \pm 0.17$& $\bf 0.001^{+0.001}_{-0.000}$& $0.0002^{+0.0002}_{-0.0001}$& $0.001$\\
        NIRSPEC + MIRI (i) &   &  $0.19$&  $\bf 0.20$& $\bf 1498.77^{+0.84}_{-1.5}$& $0.18$& $0.001$\\
    \end{tabular}
    \newline 
    \vspace{-0.5pt}
    \vspace{-8.362pt}
    
    \end{table*}
\endgroup

\subsection{The most viable of the viable extinctions}
We have fit three distinct and physically motivated models to each of the spectral datasets in the first 14 days following the BOAT.
Each of these model fits is treated independently, however, any solution to a single epoch should be consistent via the expected evolution of a GRB afterglow with the other epochs to be considered possible.
In Table\,\ref{tab:XSe1_fit}, model fit solutions that return a spectral index in the range $0.2 \leq \beta \leq 0.9$ are considered `viable', where this range is the expected viable range of spectral index at optical and NIR throughout these epochs, given the observational results of contemporary afterglow data \citep[e.g.,][]{laskar2023, levan2023, shrestha2023, sanchez-ramirez2024}.

We cannot instantly distinguish between these sets of solutions, however, we note that for $A_V = [3.5, 4.0, 4.5, 5.0]$, each epoch has at least one viable solution that indicates a self-consistent evolution of the spectral parameters may be within this extinction range.
We point out that $A_V = 7.0$ also shows that SBPL1 has viable solutions at all epochs, however, closer inspection of the break wavelengths indicates that for all X-Shooter epochs the fit component of the model spectrum to the data is well within the $\beta = -1/3$ regime.
As such, the apparent viability is driven here by the combination of the curvature due to the high extinction, and the synchrotron tail at $F(\lambda>\lambda_{\beta_1})\propto\lambda^{-1/3}$ i.e., the fit value of $\beta_1$ and $\lambda_{\beta_1}$ are well below the data range.
We therefore do not consider $A_V = 7.0$ to be viable.

\begin{figure*}
    \includegraphics[width=\textwidth]{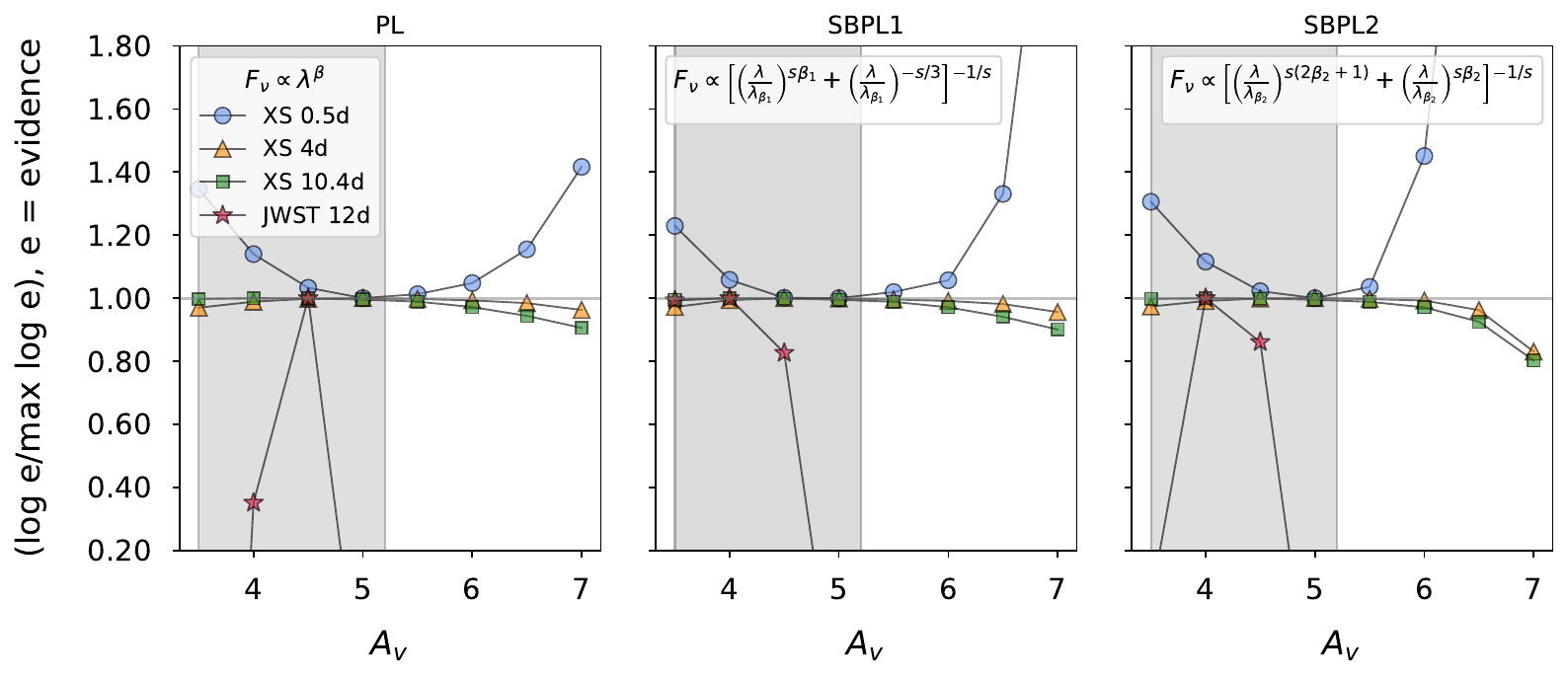}
    \includegraphics[width=\textwidth]{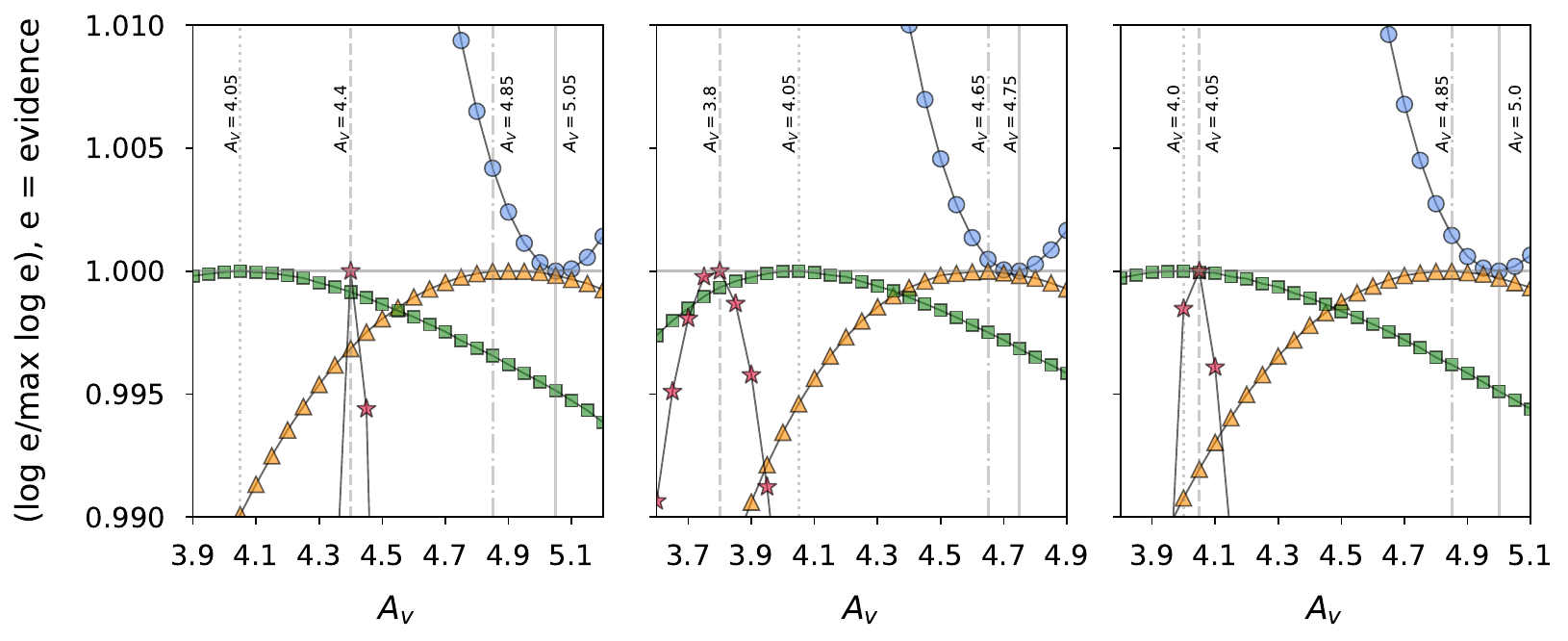}
    \caption{The ratio of the Bayesian log evidence vs the maximum log evidence for each of the spectral models, power-law, and the two smoothly broken power laws where the ratio of the models evidence with the maximum model evidence $\rightarrow 1$, then this value of $A_V$ is preferred by the data. 
    \textbf{Top panel:} The range of $A_V$ goes from 3.5 - 7.0 in 0.5 intervals. The gray shaded band is the approximate area where all four datasets have a maximum log evidence ratio of 1. 
    \textbf{Bottom panel:} Zoomed in version from the gray shaded band of the above panel, looking at an $A_V$ range from 3.50 - 5.20 in 0.05 intervals. This allows us to see more clearly where the evidence ratios converge to 1 across the datasets, which is depicted via the gray vertical lines. The values are tabulated in Table ~\ref{tab:extinction_evidence}.}
    \label{fig:log_e}
\end{figure*}

Beyond selecting viable fits, the statistically best fitting extinction value for each model and each epoch can be found by considering the Bayesian evidence.
In Figure \ref{fig:log_e}, the top panels show the ratio of the log evidence from the fits for each model at a fixed extinction, to the maximum log evidence for that set of model fits.
Where the ratio goes to unity aligns with the highest evidence extinction value for each of the models.
The X-Shooter epoch ii and iii fits have the smallest evidence at all extinctions -- as seen by the small variation in the ratio for extinction.
However, each model still has a single and distinct highest evidence case for each epoch.

\begin{figure*}
    \includegraphics[width=0.7\textwidth]{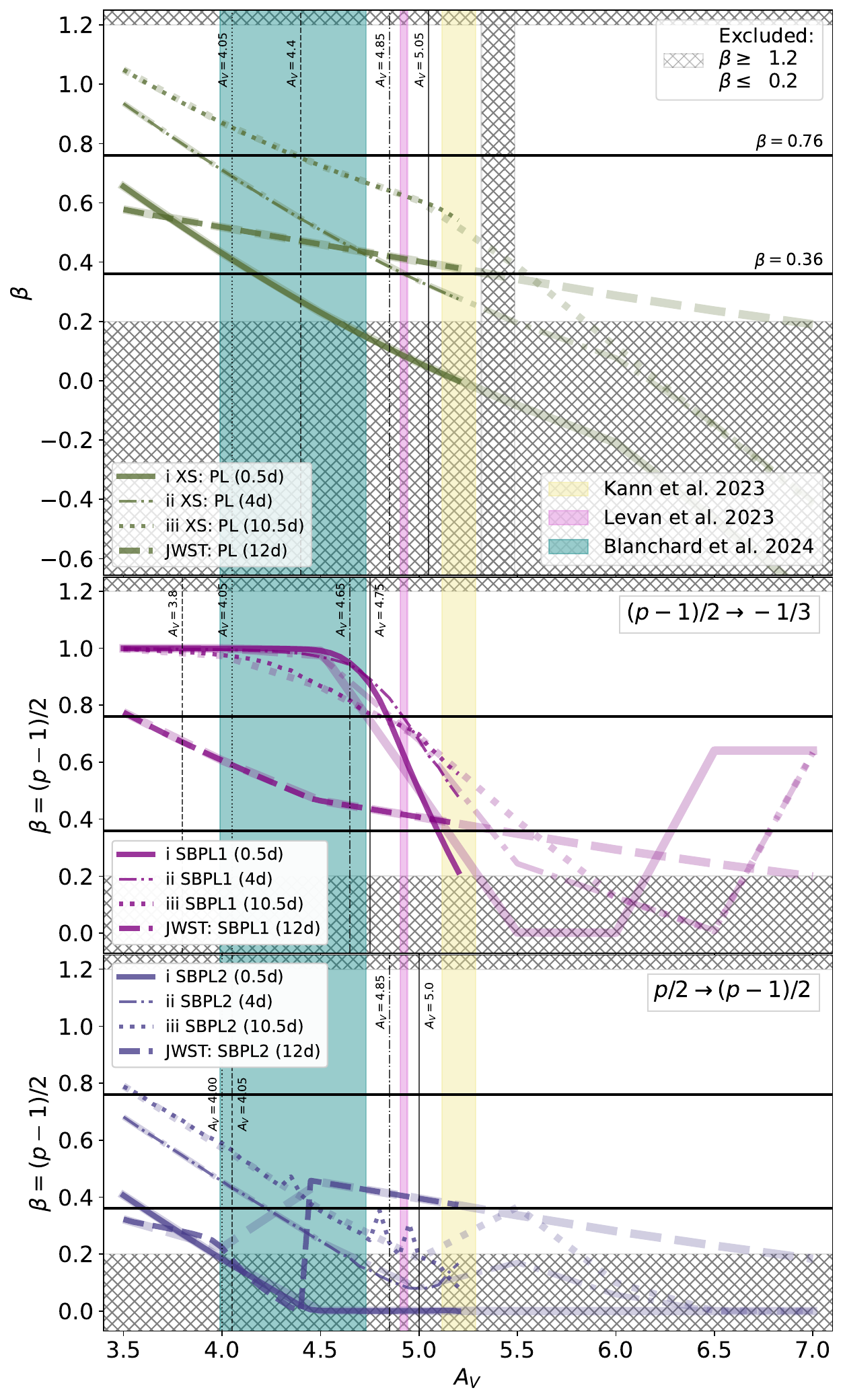}
    \vspace{-4.14pt}
    \caption{The power-law, and smoothly broken power-law fits to the 0.5, 4.0, and 10.5 day X-Shooter and the 12 day JWST spectra where $A_V$ is fixed, and $E(B-V)$ fit in the range $0.9$ to $2.2$ based on the observed spectra at 0.5 days. For the smoothly broken power-law cases, the y-axis refers to $\beta = (p-1)/2$. The shaded regions indicate the various estimates for the dominant Milky-Way line-of-sight extinction from afterglow analysis \citep{kann2023, levan2023, blanchard2024}. The more transparent lines underneath are the fitted values from $A_V$ of 3.5 - 7.0 in 0.5 intervals, with the more solid lines covering a more resolved range from $A_V$ of 3.50 - 5.20 in 0.05 intervals. The vertical lines are data-preferred values of $A_V$ inferred from Figure \ref{fig:log_e} and are styled accordingly to their epoch's formatting. {\bf Top; middle; bottom:} The three rows show the single power law (Eq.\,\ref{eq:simp}), the first SBPL (Eq.\,\ref{eq:low}), and the second SBPL (Eq.\,\ref{eq:high}) respectively. The grey hashed area indicates spectral indices that are outside of the usual fading afterglow phase, $0.2\leq\beta\leq1.2$ \citep{kann2010}. The solid horizontal lines show the highest and lowest inferred values of the spectral index found by \citet{laskar2023} and \citet{levan2023} respectively.}
    \label{fig:AV_xb}
\end{figure*}

The lower panels in Figure \ref{fig:log_e} show a zoom-in of the top panel's analysis, using a finer resolution in fixed $A_V$ values across the refined range.
The extinction values with the highest evidence for each model and at each epoch are highlighted via a vertical line (consistent with the line-style used for each epoch in Figure \ref{fig:AV_xb}).
We tabulate the highest evidence extinction for each epoch and model in Table \ref{tab:extinction_evidence}.
The extinction can be broadly considered to fall in the range, $3.80 < A_V < 5.05$, and covers the range of literature published values.

\begin{table}
    \centering
    \caption{The extinction values that correspond to the highest evidence (see, Figure \ref{fig:log_e}) from the fits of each model to the data at a given epoch. The line-of-sight extinction, $A_V$, is fixed for each iteration of the model/epoch fits. The highest evidence extinction ranges between $3.8\leq A_V \leq 5.05$ across all cases.}
    \label{tab:extinction_evidence}
    \begin{tabular}{c|c|c|c}
    Model     &  PL & SBPL1 & SBPL2 \\
    \hline
    epoch i     & 5.05 & 4.75 & 5.0 \\
    epoch ii & 4.85 & 4.65 & 4.85 \\
    epoch iii & 4.05 & 4.05 & 4.0 \\
    JWST & 4.4 & 3.8 & 4.05 \\
    \end{tabular}
    
\end{table}

\section{Evidence For Most Preferred Afterglow Models}

We examine the preferences of afterglow models employed across the viable extinction range for the JWST data (Table~\ref{tab:bayes}) by subtracting, at each $A_V$ intervals, their model's evidence scores to each other, and highlighting where the scores are highest as the most preferred one.
The same is done in Table~\ref{tab:other_bayes} for the three X-Shooter epochs, but only for $A_V = 4.0$, as it is the most preferred extinction as shown by the JWST dataset, and due to the lack of quality of the X-Shooter data themselves.

\begin{table*}
    \centering
\caption{Log evidence table comparison between different power law models for JWST. The bolded entries show the model comparisons within the column/row with the highest evidence difference, indicating which model is most heavily preferred.}
\label{tab:bayes}
    
    \resizebox{\textwidth}{!}{\begin{tabular}{|l|c|lllc|c|c|c|c|c|c|c|c|}\hline
  & & \multicolumn{3}{c}{$A_v = 3.5$}& \multicolumn{3}{|c|}{$\bf A_V = 4.0$}& \multicolumn{3}{|c|}{$A_V = 4.5$}& \multicolumn{3}{|c|}{$A_V = 5.0$}\\\hline
        
         &\textbf{Evidence (a - b)} & PL(a)& SBPL1(a)&SBPL2(a)& PL(a)& \textbf{SBPL1(a)}& SBPL2(a)& PL(a)& SBPL1(a)& SBPL2(a)& PL(a)& SBPL1(a)& {SBPL2(a)}\\\hline
 & PL(b)& 0& 11852.2& 8495.8& 9379.6& $\bf 11878.4$ & 11703.1& 11278.9& 10552.4& 10643.4& 7355.7& 4182.6&3419.1\\
 $A_v = 3.5$& SBPL1(b)& -11852.2& 0& -3356.3& -2472.5& $\bf 26.2$ & -149.1& -573.2& -1299.7& -1208.7& -4496.4& -7669.5&-8433.0\\
 & SBPL2(b)& -8495.8& 3356.3& 0& 883.8& $\bf 3382.6$ & 3207.2& 2783.1& 2056.6& 2147.6& -1140.1& -4313.2&-5076.7\\\hline
         
         &PL(b) & -9379.6& 2472.5&-883.8& 0& $\bf 2498.8$& 2323.4& 1899.3& 1172.8& 1263.8& -2023.9& -5197.0& $ -5960.5$\\
  $\bf A_V = 4.0$&\textbf{SBPL1(b)} & $\bf -11878.4$& $\bf -26.2$&$\bf -3382.6$& $\bf -2498.8$ & $\bf 0$& $\bf -175.4$& $\bf -599.5$& $\bf -1326.0$& $\bf -1235.0$& $\bf -4522.7$& $\bf -7695.8$& $\bf -8459.3$\\
         &SBPL2(b) & -11703.0& 149.1&-3207.2& -2323.4& $\bf 175.4$& 0 & -424.1& -1150.6& -1059.6& -4347.3& -7520.4& $ -8283.9$\\\hline
         
         &PL(b) & -11278.9& 573.2&-2783.1& -1899.3& $\bf 599.5$& 424.1& 0& -726.5& -635.5& -3923.2& -7096.3& $ -7859.8$\\
  $A_V = 4.5 $ &SBPL1(b) & -10552.4& 1299.7&-2056.6&  -1172.8& $\bf 1326.0$& 1150.6& 726.5& 0& 91.0& -3196.7&  -6369.8& $ -7133.3$ \\
         &SBPL2(b) & -10643.4& 1208.7&-2147.6& -1263.8& $\bf 1235.0$& 1059.6& 635.5& -91.0& 0& 3287.7& -6460.8& $  -7224.3$\\\hline
         
         &PL(b) & -7355.7& 4496.4&1140.1& 2023.9& $\bf 4522.7$& 4347.3& 3923.2& 3196.7& 3287.7& 0& -3173.1& $ -3936.6$\\
  $ A_V = 5.0$&SBPL1(b) & -4182.6& 7669.5&4313.2& 5197.0& $\bf 7695.8$& 7520.4& 7096.3& 6369.8& 6460.8&  3173.1& 0& $ -763.5$\\
         &{SBPL2(b)} & -3419.1& 8433.0&5076.7& $5960.5$& $\bf8459.3$& $8283.9$& $7859.8$& $7133.3$& $7224.3$& $3936.6$& $763.5$ & $ 0 $\\ \hline

    \end{tabular}}
\end{table*}

\begin{table*}
    \centering
\caption{Log evidence table comparison between different power law models for X-Shooter epoch I - shown that in Figure~\ref{fig:log_e}, the intrinsic spectra of X-Shooter itself is not good quality-wise hence leading to a degeneracy between the extinction and the power-law afterglow (very steep curve of $\beta$) - JWST shown that $A_V = 4.0$ is heavily preferred and so we will edit this X-Shooter table to only include $A_V = 4.0$ for all epochs. Bolded values show the highest overall log evidence factor for the entire epoch's model selection.}
\label{tab:other_bayes}
    \begin{tabular}{lc|c|c|c|}\hline
        
          &\textbf{$\bf A_V = 4.0$ // Evidence (a - b)}& PL(a)& SBPL1(a)& SBPL2(a)\\\hline
         
          &PL(b)& 0& $\bf 4380.05$& 1257.03\\
   epoch i &SBPL1(b)&  -4380.05& 0& -3123.01\\
          &SBPL2(b)& -1257.03& 3123.01&0\\ \hline
 & PL(b)& 0& $\bf 75.34$&28.33\\
 epoch ii & SBPL1(b)& -75.34& 0&-47.01\\
 & SBPL2(b)& -28.33& 47.01&0\\ \hline
 & PL(b)& 0& \bf 4.13&3.78\\
 epoch iii & SBPL1(b)& -4.13& 0&-0.35\\
 & SBPL2(b)& -3.78& $0.35$&0\\ \hline

    \end{tabular}
\end{table*}

\section{Derivation of the Double-Smoothly-Broken Power-Law Model}\label{appd:C}

Since we are trying to create an afterglow model that encompasses both the synchrotron peak and the cooling break, we will rewrite their respective smoothly broken power laws (Eq. \ref{eq:low} \& Eq. \ref{eq:high}) and their corresponding spectral indices in an alternate form. We let $X_i = \frac{\lambda}{\lambda_i}$, where $i$ can either be the synchrotron peak ($m$) or cooling break ($c$) notations,
\begin{equation}
    F_m = f_m \left[ X_m^{-\beta_1 s_1} + X_m^{-\beta_2 s_1} \right]^{-\frac{1}{s_1}},
    \label{eq:low_alt}
\end{equation}
\begin{equation}
    F_c = f_c \left[ X_c^{-\beta_2 s_2} + X_c^{-\beta_3 s_2} \right]^{-\frac{1}{s_2}},
    \label{eq:high_alt}
\end{equation}
where the 'sharpness' parameters are defined by 
\citet{granot2002}. 
$\beta_1$ describes the slope where $\lambda > \lambda_m$, $\beta_2$ is where $\lambda_m > \lambda > \lambda_c$, and $\beta_3$ lies past the cooling break at $\lambda_c > \lambda$, and $f_m$ and $f_c$ are the respective flux constants.

The overlapping segment between the two spectral breaks is where we bridge the two power laws together, this means we can equate how their fluxes would vary as $f_mX_m^{\beta_2}  =  f_cX_c^{\beta_2}$. 
Rearranging for $f_c$ and substituting back into Eq. \ref{eq:high_alt} with the intention of recovering a form that resembles Eq. \ref{eq:low_alt},
\begin{align*}
    F_c &= f_m(X_m^{\beta_2}) X_c^{-\beta_2} \left[ X_c^{-\beta_2 s_2} + X_c^{-\beta_3 s_2} \right]^{-\frac{1}{s_2}}, \\
    &= f_m X_m^{\beta_2} \left[ \frac{X_c^{-\beta_2 s_2} + X_c^{-\beta_3 s_2}}{X_c^{-\beta_2 s_2}} \right]^{-\frac{1}{s_2}}, \\
    &= f_m X_m^{\beta_2} \left[1 + X_c^{-s_2(\beta_3 - \beta_2)} \right]^{-\frac{1}{s_2}},
\end{align*}
this helps us obtain the following result,
\begin{equation}
    F_c = f_m \left[ X_m^{-\beta_2s_1}  \left[ 1 + X_c^{-s_2(\beta_3 - \beta_2)} \right]^{\frac{s_1}{s_2}} \right]^{-\frac{1}{s_1}}.
    \label{eq:fc_redone}
\end{equation}

By noticing that the entire term within the larger square bracket of Eq. \ref{eq:fc_redone} is able to describe the curve of the spectra where a cooling break occurs, we can then stitch this term within the expression that only describes the synchrotron peak in Eq. \ref{eq:low_alt}. Hence, this would form a complete solution that includes both spectral breaks and their subsequent behaviours. Knowing also that $\beta_1 = 1/3$, $\beta_2 = \beta$ \& $\beta_3 = \beta_2 + 1/2$, we therefore recover the final form used for this study of the double-smoothly-broken power-law (DSBPL) model defined by Eq. \ref{eq:DSBPL} in section \ref{sec:combined_fit}.

\bsp	
\label{lastpage}
\end{document}